\documentclass[a4paper,12pt]{article}
\usepackage[breaklinks,bookmarks]{hyperref}
\usepackage{fullpage,amsmath,amssymb,amsthm,citesort,bbold}
\usepackage[noadjust]{cite}

\makeatletter
\renewcommand\section{\@startsection{section}%
  {1}%
  {\z@}%
  {-3.5ex \@plus -1ex \@minus -.2ex}%
  {2.3ex \@plus.2ex}%
  {\normalfont\large\bfseries}}
\renewcommand\subsection{\@startsection{subsection}%
  {2}%
  {\z@}%
  {-3.25ex\@plus -1ex \@minus -.2ex}%
  {1.5ex \@plus .2ex}%
  {\normalfont\normalsize\bfseries}}
\renewcommand{\@seccntformat}[1]{{\csname the#1\endcsname}.\ \ }

\makeatother

\newenvironment{shortindent}[1]
  {\begin{list}{}%
   {\settowidth{\labelwidth}{#1}\leftmargin\labelwidth
    \advance\leftmargin by\labelsep\parsep0pt\topsep0pt}
    \sloppy\clubpenalty4000\widowpenalty4000
    \renewcommand{\makelabel}[1]{##1\hfil}}%
  {\end{list}}

\newenvironment{bulletlist}
  {\begin{list}{}%
   {\settowidth{\labelwidth}{$\bullet$\ }\leftmargin\labelwidth
    \advance\leftmargin by\labelsep\parsep0pt\topsep0pt\itemsep0pt}
    \sloppy\clubpenalty4000\widowpenalty4000
    \renewcommand{\makelabel}[1]{$\bullet$\ \hfil}}%
  {\end{list}}

\newtheoremstyle{default}
  {\topsep}
  {\topsep}
  {\itemsep0pt}
  {}
  {\bfseries}
  {.}
  {0.5em}
  {}
\newtheoremstyle{defaultit}%
  {\topsep}%
  {\topsep}%
  {\itshape\itemsep0pt}%
  {}%
  {\bfseries}%
  {.}%
  {0.5em}%
  {}%
\newtheoremstyle{clmstyle}%
  {\topsep}%
  {\topsep}%
  {\itshape\itemsep0pt}%
  {}%
  {\itshape}%
  {.}%
  {0.5em}%
  {}%
\newcommand{\proofend}{\hspace*{\fill}{$\Box$}}
\renewenvironment{proof}[1][\proofname]{\par
  \normalfont
  \topsep6pt plus6pt \trivlist
  \item[\hskip\labelsep\itshape#1.]\ignorespaces
}{%
  \proofend\endtrivlist
}
\theoremstyle{defaultit}
\newtheorem{theorem}{Theorem}
\newtheorem*{theorem*}{Theorem}

\newtheorem{proposition}{Proposition}
\newtheorem{lemma}{Lemma}
\newtheorem*{lemma*}{Lemma}
\newtheorem*{fact*}{Fact}
\newtheorem*{remark*}{Remark}
\newtheorem*{remarks*}{Remarks}
\newtheorem{fact}{Fact}

\theoremstyle{clmstyle}
\newtheorem*{claim*}{Claim}
\newtheorem*{observations*}{Observations}
\newtheorem{claim}{Claim}
\theoremstyle{default}
\newtheorem{definition}{Definition}

\renewcommand{\epsilon}{\varepsilon}
\renewcommand{\phi}{\varphi}
\renewcommand{\Re}{\operatorname{Re}}
\renewcommand{\Im}{\operatorname{Im}}
\newcommand{\C}{\mathbb{C}}
\newcommand{\D}{\mathcal{D}}
\renewcommand{\H}{\mathcal{H}}
\newcommand{\K}{\mathcal{K}}

\renewcommand{\O}{\mathcal{O}}
\newcommand{\IN}{\mathbb{N}}
\newcommand{\ZZ}{\mathbb{Z}}
\renewcommand{\k}[1]{\mathopen|#1\mathclose\rangle}
\renewcommand{\b}[1]{\mathopen\langle#1\mathclose|}
\newcommand{\bk}[2]{\mathopen\langle#1\,|\,#2\mathclose\rangle}
\newcommand{\floor}[1]{{\left\lfloor #1\right\rfloor}}

\DeclareMathOperator{\WS}{\mbox{$\mathrm{WS}$}}
\DeclareMathOperator{\MWS}{\mbox{$\mathrm{MWS}$}}
\DeclareMathOperator{\IND}{\mbox{$\mathrm{IND}$}}
\DeclareMathOperator{\DISJ}{\mbox{$\mathrm{DISJ}$}}
\DeclareMathOperator{\ND}{\mbox{$\mathrm{ND}$}}
\DeclareMathOperator{\AND}{\mbox{$\mathrm{AND}$}}
\DeclareMathOperator{\XOR}{\mbox{$\mathrm{XOR}$}}
\DeclareMathOperator{\IC}{\mbox{$\mathrm{IC}$}}
\DeclareMathOperator{\lbl}{\rm label}
\DeclareMathOperator{\var}{\rm var}
\DeclareMathOperator{\tr}{\rm tr}

\newcommand{\fn}[1]{\mathcal{#1}}

\begin{document}

\title{\vspace*{-\baselineskip}\Large\bfseries Quantum vs.\@ Classical\\[2pt]
Read-Once Branching Programs}
\author{\normalsize Martin Sauerhoff\thanks{Supported by DFG grant Sa 1053/1-1.}\\
\normalsize Universit\"at Dortmund, FB Informatik, LS 2, 44221 Dortmund, Germany.\\
{\tt\normalsize martin.sauerhoff@udo.edu}}
\date{}
\maketitle

\noindent\textbf{Abstract.}\ \ \sloppy
The paper presents the first nontrivial upper and lower bounds for (non-oblivious) quantum 
read-once branching programs. It is shown that the computational power of 
quantum and classical read-once branching programs is incomparable in the following sense:\\
(i) A simple, explicit boolean function on $2n$ input bits is presented that is 
computable by error-free quantum read-once branching programs of size $O\bigl(n^3\bigr)$, 
while each classical randomized read-once branching program and each quantum OBDD
for this function with bounded two-sided error requires size~$2^{\Omega(n)}$.\\
(ii) Quantum branching programs reading each input variable \emph{exactly once}
are shown to require size~$2^{\Omega(n)}$ for computing the set-disjointness 
function~$\DISJ_n$ from communication complexity theory with
two-sided error bounded by a constant smaller than $1/2-2\sqrt{3}/7$.
This function is trivially computable even by deterministic OBDDs of linear size.\\
The technically most involved part is the proof of the lower bound in~(ii). 
For this, a new model of quantum multi-partition 
communication protocols is introduced and a suitable extension of the information cost 
technique of Jain, Radhakrishnan, and Sen (2003) to this model is presented.

\section{Introduction}

This paper deals with the space complexity of sequential, nonuniform quantum algorithms, modeled by
quantum branching programs. It follows the general plan of developing lower bound techniques
for gradually less restricted variants of the model. This line of research is well 
motivated by the fact that, in the classical case, it has already led to practically meaningful 
time-space tradeoff lower bounds for general randomized branching programs solving
decision problems~\cite{Ajt99b,Ajt02a,Bea03a}.

Lower bounds and separation results generally come in two main flavors: 
results for multi-output-bit functions and for single-output-bit functions or decision problems. 
Of the former type are recent time-space tradeoffs for quantum circuits 
computing some practically important functions, including 
sorting~\cite{Kla03a,Aar04a,Kla04a} and boolean matrix-vector
and matrix-matrix multiplication~\cite{Kla04a,Kla04c}.

\goodbreak
Here we are concerned with lower bounds and separation results for decision problems, which
are usually harder to obtain than for multi-output-bit problems in the same model.
Such results have been proved for the uniform model of quantum finite automata (QFAs, see, e.\,g.,~\cite{Kon97,Moo00a,Amb98}). 
On the nonuniform side, general quantum branching programs and quantum OBDDs (ordered binary
decision diagrams) have been considered (see the next section for an introduction of these models).
Extending independently obtained results by \v{S}palek~\cite{Spa02a}, 
it has been shown in~\cite{Sa05a} that the logarithm of the size of general
quantum branching programs captures the space complexity of nonuniform quantum Turing machines. 
Ablayev, Moore, and Pollett~\cite{Abl02a} have proved 
that~$\text{NC}^1$ is included in the class of functions that can be 
exactly computed by quantum oblivious width-$2$ branching programs of polynomial size, 
in contrast to the classical case where width~$5$ is necessary
unless $\text{NC}^1 = \text{ACC}$.  Furthermore, exponential gaps have been established between 
the width of quantum OBDDs and classical deterministic OBDDs (Ablayev, Gainutdinova, and Karpinski~\cite{Abl01a})
and classical  randomized OBDDs, resp. (Na\-ka\-ni\-shi, Ha\-ma\-gu\-chi, and Ka\-shi\-wa\-ba\-ra~\cite{Nak00}).
Finally, it has been shown in~\cite{Sa05a} that the classes of functions with polynomial size quantum OBDDs 
and deterministic OBDDs are incomparable and an example of a partially defined function for which quantum OBDDs
are exponentially smaller than classical randomized ones has been presented.

Proving lower bounds on the space complexity of quantum algorithms for models that are more general than QFAs or 
quantum OBDDs and solve explicit decision problems has been open so far. In particular, previous results in this context 
have been limited to models that are oblivious, i.\,e., are required to read their input bits in a fixed order. 
Here we consider the non-oblivious model of quantum read-once branching programs, which are quantum branching programs 
that during each computation may access each input bit at most once.
The logarithm of the size of quantum read-once branching programs is a lower bound on the space-complexity 
of (uniform or nonuniform) quantum read-once Turing machines. This follows by an easy adaptation of the proof in~\cite{Sa05a} 
for general quantum branching programs. On the other hand, all upper bounds presented
here in terms of quantum read-once branching programs can easily be modified to work also 
for (uniform or nonuniform) quantum read-once Turing machines. 

We prove the first nontrivial upper and lower bounds for quantum read-once branching
programs. As our first main result, we present a simple function for which
quantum read-once branching programs are exponentially smaller than 
classical randomized ones. This result is even for a total function 
(compare this to the fact that analogous results for quantum OBDDs~\cite{Sa05a} and
quantum one-way communication complexity~\cite{Yos04b} known so far are only for partially 
defined functions). We use the \emph{weighted sum function} 
due to Savick\'{y} and \v{Z}\'{a}k~\cite{Sav00} as a building block.
For a positive integer~$n$ and $x=(x_1,\ldots,x_n)\in\{0,1\}^n$, let $p(n)$ be the smallest prime
larger than~$n$ and let $s_n(x) = \bigl(\sum_{i=1}^n i\cdot x_i\bigr)\bmod p(n)$. Define the weighted sum
function by $\WS_n(x) = x_{s_n(x)}$ if $s_n(x)\in\{1,\ldots,n\}$ and~$0$ otherwise. 
For a further input vector $y = (y_1,\ldots,y_n)\in\{0,1\}^n$ define the \emph{mixed weighted sum function} 
by $\MWS_n(x,y) = x_i\oplus y_i$ if $i = s_n(x) = s_n(y)\in\{1,\ldots,n\}$ and~$0$ otherwise. 

{\samepage
\begin{theorem}\label{the:mws}\sloppy\hbadness=5200
Each randomized read-once branching program and each quantum OBDD computing~$\MWS_n$ with two-sided error 
bounded by an arbitrary constant smaller than~$1/2$ requires size~$2^{\Omega(n)}$, 
while~$\MWS_n$ can be computed by an error-free quantum read-once branching
program of size $O\bigl(n^3\bigr)$.
\end{theorem}

}

\goodbreak
The above result shows that being able to choose different variable orders 
for different inputs may help a lot for quantum read-once algorithms, even compared to classical randomized read-once algorithms that
are allowed the same option. On the other hand, combining the read-once property with the usual 
unitarity constraint for quantum algorithms (required by physics) can also turn out to be
a severe restriction on the computing power. It has already been shown in~\cite{Sa05a} that 
quantum OBDDs for the set-disjointness function $\DISJ_n$ from communication complexity theory, 
defined by $\DISJ_n(x,y) = {\neg(x_1y_1\lor\cdots\lor x_ny_n)}$ 
for $x=(x_1,\ldots,x_n),y=(y_1,\ldots,y_n)\in\{0,1\}^n$, require size $2^{\Omega(n)}$. 
As our second main result, we prove a lower bound of the same order
even for the non-oblivious case.
We need the additional assumption here that the branching programs do not only read each
input variable at most once, but even \emph{exactly} once.

\begin{theorem}\label{the:disj}\sloppy
Each quantum branching program that reads each input variable exactly once and computes~$\DISJ_n$ with 
two-sided error bounded by a constant smaller than~$1/2 - 2\sqrt{3}/7$ $($$\approx 0.005$$)$
has size~$2^{\Omega(n)}$.
\end{theorem}

Note that $\DISJ_n$ can be trivially computed by deterministic OBDDs of linear size.
With the usual ``uncomputing'' trick it is also easy to construct
a reversible (and thus quantum) oblivious read-twice branching program of 
linear size for this function.

The proof of the above lower bound is considerably more involved and uses a more advanced technique 
than that for quantum OBDDs in~\cite{Sa05a}, although both rely on arguments from information theory. 
We use the general information-theoretical framework that Bar-Yossef, Jayram, Kumar,
and Sivakumar~\cite{Yos04a} have developed for classical randomized communication complexity 
and that they have applied, among other results, for an elegant new proof of a linear lower bound 
for the disjointness function. 
Furthermore, we exploit main ideas from the recent extension to the quantum case for 
a bounded number of rounds due to Jain, Radhakrishnan, and Sen~\cite{Jai03a,Jai03b},
who in turn relied on technical tools due to
Klauck, Nayak, Ta-Shma, and Zuckerman~\cite{Kla01b,Kla04b}.
For formalizing the proof, we introduce a new model of
\emph{quantum (one-way) multi-partition protocols} that allows protocols to use more 
than one input partition and may be interesting for its own sake. 
(See~\cite{Sa04b} for a nondeterministic, classical variant of this model.)
The core part of the proof is a lower bound of $\Omega(1)$ on the information cost of 
quantum multi-partition protocols computing the AND of two bits.
This complements a similar bound due to Jain, Radhakrishnan, and Sen that only 
works for a single input partition, but for any constant 
number of rounds instead of only one round here.

It remains open whether the lower bound in Theorem~\ref{the:disj} remains true for quantum 
read-once branching programs that are not forced to read each variable at least once during
any computation. It is easy to enforce this property for classical read-once 
branching programs while maintaining polynomial size, but it is not clear how 
to do this in the quantum case due to the required unidirectionality of
the programs (see the next section).

The rest of the paper is organized in the obvious way: In the next section,
we define the variants of quantum branching programs considered here.
In two further sections, we present the proofs of the main results.

\section{Preliminaries}\label{sec:prelim}

{\sloppy
We assume a general background on quantum computing and quantum information theory
(as provided, e.\,g., by the textbook of Nielsen and Chuang~\cite{Nie00}) and on
classical branching programs (BPs) (see, e.\,g., the textbook of Wegener~\cite{Weg00}). 
We start with the definition of general quantum branching programs.

}

\begin{definition}\label{def:qbp}\sloppy
A \emph{quantum branching program (QBP) over the variable set
$X = \{ x_1,\ldots,x_n\}$} is a directed multigraph $G = (V,E)$
with a \emph{start node} $s\in V$ and a set $F\subseteq V$ of sinks.
Each node $v\in V-F$ is labeled by a variable $x_i\in X$ and
we define $\var(v) = i$. Each node $v\in F$ carries a label from $\{0,1\}$, denoted by $\lbl(v)$.
Each edge $(v,w)\in E$ is labeled by a boolean constant $b\in\{0,1\}$ and a
\emph{(transition) amplitude}~$\delta(v,w,b)\in\C$.
We assume that there is at most one edge carrying the same boolean label
between a pair of nodes and set $\delta(v,w,b) = 0$ 
for all $(v,w)\not\in E$ and $b\in\{0,1\}$. 

The graph~$G$ is required to satisfy the following two constraints.
First, it has to be \emph{well-formed}, meaning that for each pair of 
nodes $u,v\in V-F$ and all assignments $a=(a_1,\ldots,a_n)$ to the variables in~$X$,
$\sum_{w\in V} \delta^*(u,w,a_{\var(u)})\delta(v,w,a_{\var(v)}) = 1$ if $u=v$ and~$0$ otherwise.
Second, $G$ has to be \emph{unidirectional}, which means that for each $w\in V$, 
all nodes $v\in V$ such that $\delta(v,w,b)\neq 0$ for some $b\in\{0,1\}$ are labeled by
the same variable.

A \emph{computational state} of the QBP is a pure quantum state over the Hilbert
space $\H = \C^{|V|}$ spanned by an ON-basis $(\k{v})_{v\in V}$.
The computation for an input $a=(a_1,\ldots,a_n)$ starts with
the computational state~$\k{s}$, called \emph{initial state}. 
Let the QBP be in the computational state~$\k{\psi} = \sum_{v\in V} \alpha_v\k{v}\in\H$ 
at the beginning of a computation step. Then the QBP first carries out a projective measurement
of the output label at the nodes in~$\k{\psi}$. This yields the result~$r\in\{0,1\}$ with
probability $\sum_{v\in F,\,\lbl(v)=r} |\alpha_v|^2$. If one of these 
events occurs, the respective output is produced and the computation stops. 
The computation carries on for the non-sink nodes with nonzero amplitude in~$\k{\psi}$.
Let $\k{\psi'} = \sum_{v\in V-F} \alpha_v'\k{v}$ be the state obtained by projecting~$\k{\psi}$ to the 
subspace spanned by the non-sink nodes and renormalizing. Then the next computational state is defined
as $\k{\psi''} = \sum_{v\in V-F} \alpha_v'\sum_{w\in V} \delta(v,w,a_{\var(v)})\k{w}$.

The probability that \emph{$G$ outputs $r\in\{0,1\}$ on input~$a\in\{0,1\}^n$} is 
defined as the sum of the probabilities of obtaining the output~$r$ after any finite number of steps.
Let $G(a)$ be the random variable describing the output of~$G$ on input~$a$, called the 
\emph{output random variable of~$G$ for~$a$}.
We say that the function $f\colon\{0,1\}^n\to\{0,1\}$ defined on~$X$ is computed by~$G$
\begin{bulletlist}
\item with \emph{two-sided error at most~$\epsilon$}, $0\le\epsilon<1/2$, if for each~$a\in\{0,1\}^n$, 
  $\Pr\{ G(a) \neq f(a)\} \le \epsilon$; and it is computed
\item \emph{exactly} (or $G$ is an \emph{error-free} QBP for~$f$), if for each~$a\in\{0,1\}^n$,  $\Pr\{ G(a) \neq f(a) \} = 0$.
\end{bulletlist}

\goodbreak
Furthermore, by \emph{bounded two-sided error} we mean two-sided error with some 
unspecified constant bound~$\epsilon$. (Other modes of acceptance may be defined as 
usual for other quantum models of computation.)

The \emph{size} of a QBP~$G$ is the number of its nodes and is denoted by~$|G|$.
Its \emph{width} is the maximum number of nodes with the same distance from the start node.
\end{definition}

The definition of QBPs is similar to that of the uniform models of quantum finite automata (QFAs) and
quantum Turing machines (QTMs), whose relationships to the respective classical models have 
already been studied to a considerable extent (see, e.\,g., \cite{Kon97,Moo00a,Amb98,Ber97,Wat99,Wat04a}).
A strong motivation why QBPs are a natural model is provided by the fact that the logarithm
of their size and the space complexity for nonuniform QTMs are polynomially related~\cite{Spa02a,Sa05a}.
For the scenario of sublinear space bounds, it has turned out to be useful to work with \emph{unidirectional} QTMs,
i.\,e., QTMs whose directions of head movements depend only on the entered state of the finite control. 
This is the standard model in the papers of Watrous~\cite{Wat99,Wat04a} and also that used
for the simulation between QBPs and QTMs in~\cite{Spa02a,Sa05a}. The unidirectionality constraint
for QBPs (called \emph{parental condition} in~\cite{Spa02a}) turns up as a natural counterpart of that 
for QTMs required to make the simulations work. In order to prevent QBPs from being unreasonably powerful, 
it is further realistic to restrict the set of allowed amplitudes, 
see also~\cite{Sa05a}. This is no issue here, since the upper bounds
in the paper only use amplitudes from $\{0,1,\pm 1/2\}$ and
the lower bounds for QBPs are valid for arbitrary complex amplitudes.

For the construction of QBPs it is sometimes convenient to use {\em unlabeled nodes} 
with an arbitrary number of outgoing edges carrying only amplitude labels. 
An unlabeled node~$v$ can be regarded as an abbreviation for a node according to the
standard definition labeled by a dummy variable on which the considered
function does not depend. Each edge leading from the unlabeled node~$v$ to a successor~$w$ 
with amplitude~$\alpha$ is then regarded as a pair of edges from the node labeled by the
dummy variable to~$w$ that carry the boolean labels~$0$ and~$1$, resp., 
and that both have amplitude~$\alpha$. 

A special case of QBPs are \emph{reversible} classical BPs, 
where each node is reachable from at most one node~$v$ by a $0$-edge and from at most one node
$w$ by a $1$-edge and $v$ and $w$ are labeled by the same variable.
It has been proved by \v{S}palek~\cite{Spa02a} that each sequence of (possibly non-reversible) classical BPs 
with at least linear size can be simulated by a sequence of reversible ones with at most polynomial larger size.
Since randomized (general) BPs can be derandomized while maintaining polynomial size analogously to probabilistic
circuits (see~\cite{Sa99c} for details), the same is true in the randomized case.

We consider the following variants of quantum BPs defined
analogously to their classical counterparts.

\begin{definition}\label{def:restr_BPs}\item[]\vspace*{-2pt}
\begin{bulletlist}
\item A quantum BP is called \emph{leveled} if the set of its nodes can be 
  partitioned into disjoint sets $V_1,\ldots,V_{\ell}$ such that for $1\le i\le\ell-1$, 
  each edge leaving a node in~$V_i$ reaches a node in~$V_{i+1}$.
\item A \emph{quantum read-once BP} is a QBP where each variable may appear at most once on each path.
\item A {\em quantum OBDD} (quantum ordered binary decision diagram) is a quantum read-once BP 
  with an order~$\pi$ of the variables such that for each path in the graph 
  the order in which the variables appear is consistent with~$\pi$.
\end{bulletlist}
\end{definition}

\section{The Separation Result for Mixed Weighted Sum (Theorem~\ref{the:mws})}\label{sec:mws}

For the whole section, let $p = p(n)$ be the smallest prime larger than~$n$ 
for a fixed positive integer~$n$. We first deal with the easier upper bound.
Our goal is to show that $\MWS_n$ can be computed by polynomially small error-free quantum read-once BPs. 

\begin{proof}[Proof of Theorem~\ref{the:mws} -- Upper Bound]
The essence of the proof is to apply the Deutsch-Jozsa algorithm, evaluating the sums $s_n(x)$ and $s_n(y)$ 
in parallel and computing the output $x_i\oplus y_i$ if $i = s_n(x) = s_n(y)$.
We first describe the algorithm by a quantum circuit. 
We use a four-part quantum register consisting of two qubits for the Deutsch-Jozsa algorithm and two 
further parts whose basis states are indexed by $\{0,\ldots,p-1\}$.
The oracle gate for the Deutsch-Jozsa algorithm unitarily extends the
mapping~$S$ specified for $a,b\in\{0,1\}$ by 
$\k{a}\k{b}\k{0}\k{0} \mapsto {\k{a}\bigl| b\oplus (1-a)y_i \oplus a x_j\bigr\rangle\k{i}\k{j}}$,
where $i=s_n(x)$ and $j=s_n(y)$.
This gate is applied to the initial state ${(1/2)(\k{0}+\k{1})(\k{0}-\k{1})\k{0}\k{0}}$,
giving the final state $(1/2)\bigl((-1)^{y_i}\k{0}+(-1)^{x_j}\k{1}\bigr)(\k{0}-\k{1})\k{i}\k{j}$.
If a measurement of the last two parts of the quantum register yields that $i\neq j$,
the output of the circuit is~$0$ with probability~$1$. Otherwise, $i=j$ and 
measuring the first two qubits in the Hadamard basis yields the 
output~$x_i\oplus y_i = \MWS_n(x,y)$ for the first qubit with probability~$1$.

Next we describe the implementation of the obtained quantum circuit as a quantum read-once BP. 
For an easier exposition, we first use unlabeled nodes. 
We start with the construction of a subgraph~$G_S$ realizing the mapping~$S$. The nodes of~$G_S$ are
laid out on a grid with $2n+1$ rows and $4p^2$ columns, the latter labeled by $(a,b,i,j)$ 
with $a,b\in\{0,1\}$ and $i,j\in\{0,\ldots,p-1\}$. Each row represents an intermediate state
of the four-part quantum register used for the above algorithm.
The graph~$G_S$ consists of two disjoint classical reversible OBDDs $G_0$ and~$G_1$ on 
the subsets of nodes in the columns with $a = 0$ and $a = 1$, resp. We first describe how~$G_0$ works.
The computation starts at a node in row~$1$ and
column~$(0,b,0,0)$ with $b\in\{0,1\}$. The variable vector~$x$ is read (the order of the variables 
within the vector does not matter) and the node in row $n+1$ and column $(0,b,s_n(x),0)$ is reached. 
Then the variable vector~$y$ is read (again, the order of the individual variables is arbitrary) 
and the sink in row~$2n+1$ and column $(0,b\oplus y_{s_n(x)},s_n(x),s_n(y))$ is reached.
It is easy to see how the described computation can be implemented by a reversible OBDD 
with nodes on the prescribed grid.
The OBDD~$G_1$ works in the same way, but with exchanged roles of~$x$ and~$y$ and exchanged
roles of the last two column indices. Altogether, we obtain a classical reversible read-once BP
for~$G_S$ with at most $(2n+1)\cdot 4p^2$ nodes, which is of order $O\bigl(n^3\bigr)$ due to 
the prime number theorem.  

We add a new, unlabeled source that for $(a,b)\in\{0,1\}^2$
is connected to the node in row~$1$ and column $(a,b,0,0)$ of~$G_S$
by an edge with amplitude $(-1)^b(1/2)$. The sinks of~$G_S$ in row~$2n+1$ and in columns
$(a,b,i,j)$ with $i\neq j$ are replaced with $0$-sinks. All other sinks of~$G_S$ are replaced with unlabeled nodes
connected to a new level of sinks with boolean output labels. The outgoing edges of these unlabeled nodes are labeled
by amplitudes such that, together with the sinks, a measurement in the Hadamard basis is realized. 
The whole graph still has size~$O\bigl(n^3\bigr)$.

Finally, we remove the unlabeled nodes. For this, we first ensure that all nodes on the 
first level of $G_S$ are labeled by the same variable and the same for all nodes on the last 
level of~$G_S$ with variable labels. We rearrange (e.\,g.) the variable order of the OBDD~$G_1$ and update the 
OBDD accordingly. W.\,l.\,o.\,g., let $x_1$ be the first variable read in~$G_0$ and let~$y_n$ be the last. 
We move the variable~$x_1$ to the front of the variable order of~$G_1$ and $y_n$ to the end.
It is not hard to see that we can modify~$G_1$ in such a way that it complies to the new 
variable order while increasing its size by at most a constant factor and maintaining reversibility.
After this transformation, we merge the unlabeled nodes with their successors (in the case of the source)
or with their predecessors (in the case of the nodes on the level directly above the sinks). It is obvious
how the edges should be relabeled such that the resulting graph still computes the same final state
as a quantum read-once BP. We observe that after the reordering process also the unidirectionality
requirement for quantum BPs is satisfied. Altogether, we have obtained the desired quantum read-once BP 
for~$\MWS_n$ of size $O\bigl(n^3\bigr)$.
\end{proof}

Next we prove the lower bound on the size of randomized read-once BPs 
for~$\MWS_n$ with bounded error. We reuse main ideas from the proof an analogous 
lower bound for~$\WS_n$ in~\cite{Sa03a}. However, the result for $\MWS_n$ is no obvious consequence 
of that for~$\WS_n$. We have to carefully argue why, different from the quantum case, 
having two input vectors present that play the same roles does not help in the 
randomized case.

The proof employs a variant of the rectangle bound method from communication complexity theory 
(see, e.\,g., the textbook of Kushilevitz and Nisan~\cite{Kus97}) suitable for read-once BPs, 
which we fist describe. For this, we introduce some notation.
We consider boolean functions defined on the union of the disjoint 
sets of variables $X = \{x_1,\ldots,x_n\}$ and $Y = \{y_1,\ldots,y_n\}$.
For a set of variables $Z\subseteq X\cup Y$, let $2^Z$ denote the set of all assignments to~$Z$,
i.\,e., mappings from~$Z$ to~$\{0,1\}$ that we usually identify with vectors in $\{0,1\}^{|Z|}$.
A \emph{(combinatorial) rectangle} with respect to a partition~$\Pi=(\Pi_1,\Pi_2)$ of~$X\cup Y$
is a set of assignments $R = A\times B$ with $A\subseteq 2^{\Pi_1}$ and $B\subseteq 2^{\Pi_2}$.
For $\ell\in\{1,\ldots,n-1\}$ call~$R$ an \emph{$\ell$-rectangle} if $\Pi_1$ contains exactly~$\ell$
variables from~$X$ and at most~$\ell-1$ variables from~$Y$ or the same with exchanged 
roles of~$X$ and~$Y$.
Call~$R$ a \emph{one-way rectangle} if $B = 2^{\Pi_2}$. Given a function~$g$ on~$X\cup Y$,
$R$ is said to be \emph{$g$-uniform} if for all $a,a'\in A$ and $b\in B$, $g(a,b) = g(a',b)$. 

For the following, let a function~$f$ on~$X\cup Y$ and a distribution~$\D$ on the inputs of~$f$ be given.
Let $0\le\epsilon<1/2$. We describe how to prove lower bounds for deterministic read-once BPs 
whose output is allowed to differ from~$f$ on at most an $\epsilon$-fraction 
of the inputs with respect to~$\D$. By a well-known averaging argument due to Yao~\cite{Yao77}, 
this also gives lower bounds of the same size for randomized read-once BPs computing~$f$ 
with the same error probability.

The essence of the proof technique is to show that, on the one hand, any small deterministic read-once BP that 
correctly computes~$f$ on a large fraction of the inputs with respect to~$\D$ would give a rectangle with large
$\D$-measure on which~$f$ is well approximated, while on the other hand, using the specific properties
of~$f$, the $\D$-measure of any such rectangle necessarily has to be small. We now make this more 
precise. Let $R = A\times B$ be a rectangle and let $0\le\epsilon<1/2$.
A function~$g$ on~$X\cup Y$ is said to \emph{uniformly approximate~$f$ on $R$ with 
error $\epsilon$ with respect to~$\D$}, if for all $a\in A$, 
$g$ differs from~$f$ for at most an $\epsilon$-fraction of the inputs in~$\{a\}\times B$
with respect to~$\D$. The following main lemma of the proof technique is a variant of a similar
statement from~\cite{Sa03a}, where the uniform distribution and functions on a single set of variables
have been considered.

\begin{lemma}\label{lem:rect_tech}
Let $X = \{x_1,\ldots,x_n\}$ and $Y = \{y_1,\ldots,y_n\}$. 
Let $f$ be a boolean function on $X\cup Y$ and let~$\D$ be a distribution
on the inputs of~$f$. Let $\ell\in\{1,\ldots,n-1\}$ and ${0\le\epsilon<\epsilon'<1/2}$. 
Then for every deterministic read-once~BP~$G$ computing a function~$g$ that differs from~$f$ on at most
an $\epsilon$-fraction of the inputs with respect to~$\D$ there is a one-way $\ell$-rectangle~$R$ 
that is $g$-uniform, on which~$g$ uniformly approximates~$f$ 
with error at most~$\epsilon'$ with respect to~$\D$, and which satisfies $\D(R) \ge (1-\epsilon/\epsilon')/(2n|G|)$.
\end{lemma}

\begin{proof}
By an easy adaptation of the well-known proof technique of Borodin, Razborov, and Smolensky~\cite{Bor93}
(see also~\cite{Weg00}, Section~7.6), we get a partition of the input space into at most ${k\le 2n|G|}$ 
one-way $\ell$-rectangles $R_1 = A_1\times B_1,\ldots,R_k = A_k\times B_k$ that are all $g$-uniform. 
We claim that there is an~$i\in\{1,\ldots,k\}$ and a subset $A_i'\subseteq A_i$ such that
for $R = A_i'\times B_i$, $\D(R) \ge {(1-\epsilon/\epsilon')/k}$ and $g$ uniformly 
approximates~$f$ on~$R$ with error~$\epsilon'$ with respect to~$\D$. 
This obviously suffices to prove the claim.

Let $A^* = A_1\cup\cdots\cup A_k$. For each ${x\in A^*}$, let $(\Pi_1(x),\Pi_2(x))$ be the partition
of the input variables used by the rectangle to which~$x$ belongs,
and let $S_x = {\{x\}\times 2^{\Pi_2(x)}}$.
Let $A = \{ x\in A^* \mid \D(S_x) > 0 \}$.
For each~$x\in A$ let $\epsilon(x)$ be the $\D$-fraction of inputs 
from~$S_x$ for which~$g$ differs from~$f$.
Due to the definitions, the sets $S_x$, $x\in A$, are disjoint and their
union has $\D$-measure~$1$. Hence, by the law of total probability,
$\sum_{x\in A} \epsilon(x)\,\D(S_x) \le \epsilon$.
Let $A' = \mbox{$\{ x\in A \mid \epsilon(x) \le \epsilon' \}$}$
and let $S$ be the union of all $S_x$ for~$x\in A'$. By Markov's 
inequality, $\D(S) \ge 1-\epsilon/\epsilon'$.
By averaging, there is a set $A'' \subseteq A'$ such that for 
the union~$S'$ of all $S_x$ with $x\in A''$, we have $\D(S') \ge \D(S)/k$
and all inputs from~$A''$ belong to the same rectangle. 
Let~$(\Pi_1,\Pi_2)$ be the partition of input variables of this rectangle.
It is now obvious that the set $R = A''\times 2^{\Pi_2}$ with $A''\subseteq 2^{\Pi_1}$
and $\D(R) \ge (1-\epsilon/\epsilon')/k$
is a one-way $\ell$-rectangle with the desired properties.
\end{proof}

Next we cite two technical lemmas also used in~\cite{Sa03a} that build the common 
core of the lower bounds both for the mixed weighted sum function~$\MWS_n$
and the usual weighted sum function~$\WS_n$.
The first lemma allows us to argue that partial weighted sums
of enough random bits are essentially uniformly distributed over
the whole range of possible values.

\begin{lemma}[\cite{Sa03a}]\label{lem:sol}\sloppy\hbadness=2000
Let $q = q(n)$ be a sequence of primes and 
let $n\le q-1$ and $n = \Omega\bigl(q^{2/3+\delta}\bigr)$ for 
any constant \mbox{$\delta > 0$}.
Let $a_1,\ldots,a_n,b\in\ZZ_q^* = \ZZ_q-\{0\}$ 
where the numbers $a_1,\ldots,a_n$ are pairwise different. 
Then for $(x_1,\ldots,x_n)\in\{0,1\}^n$ chosen uniformly at random,
${\bigl|{\Pr\{ a_1 x_1 + \cdots + a_n x_n \equiv b \bmod q \}} - 1/q\bigr|} = 2^{-\Omega\left(q^{3\delta}\right)}$.
\end{lemma}

In the second lemma, we consider the \emph{index function}~$\IND_n\colon\{0,1\}^n\times\{1,\ldots,n\}$ from 
communication complexity theory defined for $u\in\{0,1\}^n$ and $v\in\{1,\ldots,n\}$ by $\IND_n(u,v) = u_v$.
We state an upper bound on the size of one-way rectangles on which~$\IND_n$ is well approximated that 
is implicit in a couple of papers, the earliest one being probably that of Kremer, Nisan, and Ron~\cite{Kre99}. 
For the sake of completeness, we include the easy proof. Here and in the following, 
$U$ denotes the uniform distribution on the domain implied by its respective argument.

\begin{lemma}[\cite{Kre99}]\label{lem:index}
Let $\epsilon$ be a constant with $0\le\epsilon<1/2$.
Let $R = A\times\{1,\ldots,n\}$ with $A\subseteq\{0,1\}^n$ be a one-way rectangle for which a 
function~$g$ exists such that~$R$ is $g$-uniform and~$g$ uniformly approximates~$\IND_n$ on~$R$ 
with error~$\epsilon$ with respect to~$U$. Then $U(R) = 2^{-\Omega(n)}$.
\end{lemma}

\begin{proof}
Since~$R$ is $g$-uniform, there is a vector~$r\in\{0,1\}^n$ such that, for each $a\in A$,
$(g(a,1),\ldots,g(a,n)) = r$. Since~$g$ uniformly approximates~$\IND_n$ on~$R$ with error at most~$\epsilon$
with respect to the uniform distribution, $r$ has Hamming distance at most~$\floor{\epsilon n}$ to each 
vector in~$A$. It follows that $|A|$ is upper bounded by the size of Hamming balls of radius~$\floor{\epsilon n}$,
which is known to be at most $2^{H(\epsilon)n}$, where~$H(x) = -(x\log x+(1-x)\log(1-x))$ for $x\in[0,1]$ is the 
binary entropy function. Thus, $U(R) = |R|/\bigl(n\cdot 2^n\bigr) = |A|/2^n \le 2^{-(1-H(\epsilon))n} = 2^{-\Omega(n)}$.
\end{proof}

Now we describe the details that are particular to the function~$\MWS_n$. 
For the rest of the section, let $X = \{x_1,\ldots,x_n\}$ and $Y = \{y_1,\ldots,y_n\}$ 
be the sets of variables on which~$\MWS_n$ is defined. Recall that $p = p(n)$ is 
the smallest prime larger than~$n$.
We concentrate on the set of difficult inputs $D = \{ (x,y) \mid s_n(x) = s_n(y) \}$
by working with the distribution~$\D$ with $\D(x,y) = 1/|D|$ if $(x,y)\in D$ and $\D(x,y) = 0$
otherwise. 

As a preparation of the proof of the lower bound for randomized read-once BPs
computing~$\MWS_n$, we derive some basic facts about the considered one-way rectangles.
We use the following notation. For a set $S\subseteq X$ (or ${S\subseteq Y}$) 
of variables and a partial assignment~$a$ that fixes at least all variables in~$S$, let
$\sigma_{S}(a) = {\bigl(\sum_{v\in S} i(v)\cdot a(v)\bigr)\bmod p}$, where~$i(v)\in\{1,\ldots,n\}$ 
denotes the index of the variable~$v$ in~$X$ (or~$Y$, resp.), and~$a(v)$ is the 
value that it obtains by the assignment~$a$.

\begin{lemma}\label{lem:low_level_rect_sizes}
Let $\ell = n - \Theta\bigl(p^{2/3+\delta}\bigr)$ for some constant~$\delta$ with $0 < \delta < 1/3$.
Let $\Pi=(\Pi_1,\Pi_2)$ be a partition of~$X\cup Y$ with $|\Pi_1\cap X|=\ell$ and $|\Pi_1\cap Y|\le\ell-1$.
Let $R = A\times 2^{\Pi_2}$ with $A\subseteq 2^{\Pi_1}$ and suppose there are $i_x,i_y\in\{0,\ldots,p-1\}$ such that
for all $a\in A$, $\sigma_{\Pi_1\cap X}(a) = i_x$ and $\sigma_{\Pi_1\cap Y}(a) = i_y$. 
For each $k\in\{0,\ldots,p-1\}$ define $B_k$ as the set of all assignments $b\in 2^{\Pi_2}$ with
$\sigma_{\Pi_2\cap X}(b) \equiv {(k-i_x) \bmod p}$ and $\sigma_{\Pi_2\cap Y}(b) \equiv {(k-i_y) \bmod p}$.
Then we have the following.
\begin{shortindent}{(ii)}
\item[(i)]
  For each $k\in\{0,\ldots,p-1\}$ and $(a,b)\in A\times B_k$, $\sigma_X(a,b) = \sigma_Y(a,b) = k$.
  Furthermore, $U(B_k) = (1/p^2)\cdot(1\pm o(1))$ and $\D(A\times B_k) = (1/p)\cdot U(R)\cdot (1\pm o(1))$.
\item[(ii)]
  $\D(R) = U(R)\cdot(1\pm o(1))$.
\end{shortindent}
\end{lemma}

\begin{proof}
\emph{Part~(i):} 
The first part of the statement is obvious. It remains to prove the claims about $U(B_k)$ and~$\D(A\times B_k)$.
Let $b$ denote an assignment from~$B_k$ chosen uniformly at random. Then, using that 
disjoint parts of~$b$ are independent of each other and applying Lemma~\ref{lem:sol}, we get
\begin{align*}
  U(B_k) &\ =\  \Pr\{ \sigma_{\Pi_2\cap X}(b) \equiv k-i_x \,\land\, \sigma_{\Pi_2\cap Y}(b) \equiv k-i_y \}\\
         &\ =\  \Pr\{ \sigma_{\Pi_2\cap X}(b) \equiv k-i_x\}\cdot\Pr\{ \sigma_{\Pi_2\cap Y}(b) \equiv k-i_y \}
         \ =\  \frac{1}{p^2}\cdot(1\pm o(1)).
\end{align*}
Furthermore, also by Lemma~\ref{lem:sol}, $U(D) = (1/p)\cdot(1\pm o(1))$. Again by the
independence of disjoint parts of uniformly random assignments and by
observing that ${A\times B_k\subseteq D}$ and ${U(A) = U(R)}$, we obtain
\[
  \D(A\times B_k) 
  \ =\  \frac{U((A\times B_k)\cap D)}{U(D)} 
  \ =\  \frac{U(A)\cdot U(B_k)}{U(D)}
  \ =\  \frac{1}{p}\cdot U(R)\cdot (1\pm o(1)).                 
\]

\emph{Part~(ii):} This follows from the first part, since $R\cap D$ is the disjoint 
union of the sets $A\times B_k$ over all $k=0,\ldots,p-1$.
\end{proof}

Finally, we are ready to prove the desired lower bound on the size of randomized read-once BPs for~$\MWS_n$.

\begin{proof}[Proof of Theorem~\ref{the:mws} -- Lower bound for randomized read-once BPs]\hbadness=6000\hskip0pt plus0.25em
Following the outline above, we prove the lower bound for deterministic read-once BPs
that correctly compute~$\MWS_n$ on a large fraction of the inputs. Let $0\le\epsilon_G<1/2$ be any 
constant and let~$G$ be a deterministic read-once BP computing a function~$g$ that differs from~$\MWS_n$
on at most an $\epsilon_G$-fraction of the inputs with respect to~$\D$.
Choose $\ell = n - \Theta\bigl(p^{2/3+\delta}\bigr)$ for a some constant~$\delta$ with ${0 < \delta < 1/3}$.
Let $\epsilon$ be a constant with $\epsilon_G < \epsilon < 1/2$. Let~$R$ be a one-way $\ell$-rectangle
that is $g$-uniform and on which $\MWS_n$ is uniformly approximated by~$g$ with error at most~$\epsilon$.
We prove that $\D(R) =  2^{-\Omega(n)}$. By Lemma~\ref{lem:rect_tech}, this yields the
desired lower bound $|G| = 2^{\Omega(n)}$.

Let $\Pi = (\Pi_1,\Pi_2)$ be the partition of the input variables used by~$R$, where w.\,l.\,o.\,g.\@
${|\Pi_1\cap X| = \ell}$ and ${|\Pi_1\cap Y|\le \ell-1}$. Let $R = A_R\times 2^{\Pi_2}$ with
$A_R\subseteq 2^{\Pi_1}$.  Using averaging, we fix an assignment $a\in 2^{\Pi_1\cap Y}$ and an 
$i_x\in\{0,\ldots,p-1\}$ such that for the set~$A$ of all assignments~$a'\in A_R$ 
that are consistent with~$a$ and satisfy $\sigma_{\Pi_1\cap X}(a') = i_x$, we have
$\D\bigl(A\times 2^{\Pi_2}\bigr) \ge \D(R)/\bigl(p\cdot 2^{|\Pi_1\cap Y|}\bigr)$. 
Let $i_y = \sigma_{\Pi_1\cap Y}(a)$. Let $R' = {\bigl\{ x\in A\times 2^{\Pi_2} \bigm| {\D(x) > 0} \bigr\}}$. 
Since~$g$ approximates~$\MWS_n$ uniformly on~$R$ with error at most~$\epsilon$ with respect to~$\D$, 
we know that~$g$ differs from~$\MWS_n$ for at most an $\epsilon$-fraction of the inputs in~$R'$ with respect to~$\D$.

Let $\Pi_1\cap X = \{ x_{j_1},\ldots,x_{j_\ell}\}$. 
We observe that, due to the prime number theorem, $p \le n + o(n)$ and thus $\ell \ge n - o(n)$ 
and $\ell/p \ge 1-o(1)$. Let $B_0,\ldots,B_{p-1}\subseteq 2^{\Pi_2}$ be the sets of assignments
according to Lemma~\ref{lem:low_level_rect_sizes} for $R'$ and $i_x,i_y$. 
Let $B = B_{j_1}\cup\cdots\cup B_{j_\ell}$. Then we have the following.

\setcounter{claim}{0}
\begin{claim}\label{clm:mws_i}
The function~$g$ differs from~$\MWS_n$ on at most a fraction of $\epsilon\cdot(1+o(1))$
of the inputs in~$A\times B$ with respect to the uniform distribution.
\end{claim}

\begin{proof}[Proof of Claim~\ref{clm:mws_i}]
Due to part~(i) of Lemma~\ref{lem:low_level_rect_sizes},
$\D(A\times B) \ge (\ell/p)\cdot U(R')\cdot (1-o(1)) \ge U(R')\cdot (1-o(1))$.
On the other hand, by part~(ii) of Lemma~\ref{lem:low_level_rect_sizes}, $\D(R') \le U(R')\cdot(1+o(1))$.
Thus, the inputs in~$A\times B$ cover at least a $(1-o(1))$-fraction of the rectangle~$R'$ 
with respect to~$\D$. It follows that $g$ differs from~$\MWS_n$ on at most a fraction 
of~$\epsilon\cdot(1+o(1))$ of the inputs in~$A\times B$ with respect to~$\D$. 
Since $A\times B\subseteq D$, the same is true for the uniform distribution.
\end{proof}

Next we further reduce the obtained set~$A\times B$ by picking appropriate representatives of each of
the subsets $B_{j_1},\ldots,B_{j_\ell}$ of~$B$. 

\begin{claim}\label{clm:mws_ii}
There are $b_1\in B_{j_1},\ldots,b_{\ell}\in B_{j_\ell}$ such that~$g$ differs from~$\MWS_n$ on 
at most a fraction of $\epsilon\cdot(1+o(1))$ of the inputs in $R'' = A\times\{b_1,\ldots,b_{\ell}\}$ 
with respect to the uniform distribution.
\end{claim}

\begin{proof}[Proof of Claim~\ref{clm:mws_ii}]\sloppy\hbadness=4000
We choose a collection of disjoint subsets $\{b_1,\ldots,b_{\ell}\}$ of~$B$ with ${b_1\in B_{j_1}},\ldots,b_{\ell}\in B_{j_\ell}$
whose union~$B'$ is as large as possible. Since $U(B_k) \ge (1/p^2)\cdot(1-o(1))$ for each $k=0,\ldots,p-1$
by part~(i) of Lemma~\ref{lem:low_level_rect_sizes}, we can ensure that $U(B') \ge {(\ell/p^2)\cdot(1-o(1))} \ge {(1/p)\cdot(1-o(1))}$. 
On the other hand, also by Lemma~\ref{lem:low_level_rect_sizes}, $U(B) \le (1/p)\cdot (1+o(1))$. Hence,
the set $A\times B'$ covers at least a $(1-o(1))$-fraction of the inputs in~$A\times B$.
It follows that the relative error of~$g$ on~$A\times B'$ with respect to the uniform distribution is bounded 
by some~$\epsilon'$ with~$\epsilon' \le \epsilon\cdot(1+o(1))$. By averaging, there is 
thus at least one subset~$\{b_1,\ldots,b_{\ell}\}$ in $B'$ such that $A\times\{b_1,\ldots,b_{\ell}\}$ 
has relative error~$\epsilon'$ with respect to the uniform distribution.
\end{proof}

Let $R'' = A\times\{b_1,\ldots,b_{\ell}\}$ be a rectangle according to the above claim.
Now we apply the result for the index function from Lemma~\ref{lem:index}.
For simplicity, we assume that ${j_1 = 1},\ldots,{j_{\ell} = \ell}$ such that 
the set of all restrictions of the assignments in~$A$ to the variables in~$\Pi_1\cap X$
can be identified in the obvious way with a subset $A_{\rm IND}\subseteq\{0,1\}^{\ell}$ of the same size.
Recall that for each assignment in~$A$, the variables in~$\Pi_1\cap Y$ are fixed according to the assignment~$a$ chosen above.
We regard $R_{\rm IND} = A_{\rm IND}\times\{1,\ldots,\ell\}$ as a one-way rectangle for the index function~$\IND_{\ell}$.
Define the function~$h$ on inputs $u\in\{0,1\}^{\ell}$ and $v\in\{1,\ldots,\ell\}$ by
\[
  h(u,v) \ =\ 
  \begin{cases}
  g((u,a),b_v)\oplus a(y_v),   & \text{if $y_v\in\Pi_1$; and}\\
  g((u,a),b_v)\oplus b_v(y_v), & \text{if $y_v\in\Pi_2$;}
  \end{cases}
\]
where we regard~$u$ as an assignment to~$\Pi_1\cap X$ in the argument of~$g$.
Since $b_v\in B_v$ and for each $a'\in A$, $\sigma_X(a',b_v) = \sigma_Y(a',b_v) = v$, 
\[
  \MWS_n((u,a),b_v)
  \ =\ 
  \begin{cases}
  u(x_v) \oplus a(y_v),   & \text{if $y_v\in\Pi_1$; and}\\
  u(x_v) \oplus b_v(y_v), & \text{if $y_v\in\Pi_2$;}
  \end{cases}
\]
and $h(u,v) = u_v = \IND_{\ell}(u,v)$ if $g((u,a),b_v) = \MWS_n((u,a),b_v)$.

The rectangle $R_{\rm IND}$ is $h$-uniform since $R''$ is $g$-uniform and the values~$a(y_v)$
and $b_v(y_v)$, resp., added to the output of~$g$ depend only on the second part~$v$ of the input. 
Since $g$ differs from $\MWS_n$ on at most a fraction of $\epsilon' = \epsilon\cdot(1+o(1))$ of
the inputs of~$R''$ with respect to the uniform distribution, $h$ differs from $\IND_{\ell}$ 
on at most an $\epsilon'$-fraction of~$R_{\rm IND}$ with respect to the uniform distribution.
By Lemma~\ref{lem:index}, it follows that $U(R_{\rm IND}) = 2^{-\Omega(\ell)}$. Furthermore,
\[
  U(R') \ =\ |A|/2^{|\Pi_1|} \ =\  2^{-|\Pi_1\cap Y|}\cdot |A_{\rm IND}|/2^{\ell} \ =\  2^{-|\Pi_1\cap Y|}\cdot U(R_{\rm IND})
\]
and, by part~(ii) of Lemma~\ref{lem:low_level_rect_sizes}, $\D(R') \le U(R')\cdot(1+o(1))$.
Finally, $\D(R) \le p\cdot 2^{|\Pi_1\cap Y|}\cdot \D(R')$. Putting everything together, we have shown that
$\D(R) = p\cdot 2^{-\Omega(\ell)}$.
Since $p \le n+o(n)$ and $\ell \ge n-o(n)$, this bound is of the desired size.
\end{proof}

The lower bound for quantum OBDDs stated in Theorem~\ref{the:mws} follows by standard communication
complexity arguments and the properties of~$\MWS_n$ already used above.

\begin{proof}[Proof of Theorem~\ref{the:mws} -- Lower bound for quantum OBDDs]
Let~$G$ be a quantum OBDD computing~$\MWS_n$ with error bounded by a constant~$\epsilon$, $0\le\epsilon<1/2$.
Let $\ell = n - \Theta\bigl(p^{2/3+\delta}\bigr)$ for some constant $\delta$ with $0 < \delta < 1/3$.
Appropriately cutting the list of variables used as the variable order for~$G$ in two parts 
gives a partition~$\Pi = (\Pi_1,\Pi_2)$ of the set of variables~$X\cup Y$ that, w.\,l.\,o.\,g.,
satisfies $|\Pi_1\cap X| = \ell$ and $|\Pi_1\cap Y| \le \ell-1$.
Choose $a\in 2^{\Pi_1\cap Y}$ somehow arbitrarily and
let $i_y = \sigma_{\Pi_1\cap Y}(a)$. Furthermore, again w.\,l.\,o.\,g., suppose that
$\Pi_1\cap X = \{ 1,\ldots,\ell \}$. For any $i_x\in\{0,\ldots,p-1\}$, 
Lemma~\ref{lem:sol} yields the existence of assignments $b_{i_x,1},\ldots,b_{i_x,\ell}\in 2^{\Pi_2}$ 
such that $\sigma_{\Pi_2\cap X}(b_{i_x,j}) \equiv (j-i_x) \bmod p$ and
$\sigma_{\Pi_2\cap Y}(b_{i_x,j}) \equiv (j-i_y) \bmod p$ for $j=1,\ldots,\ell$.

The given quantum OBDD~$G$ can now be used by the two players Alice and Bob in a quantum one-way communication
protocol for~$\IND_{\ell}$ as follows. Let $u\in\{0,1\}^{\ell}$ and $v\in\{1,\ldots,\ell\}$ be
the inputs for $\IND_{\ell}$. Alice follows the computation in~$G$ for the partial input $(u,a)$,
regarding~$u$ as an assignment to the variables in~$\Pi_1\cap X$, and sends the reached superposition
as well as the partial weighted sum $\sigma_{\Pi_1\cap X}(u)$ to Bob. Bob finishes the computation
of~$G$ using the partial input~$b_{i_x,v}$ and outputs the XOR of output bit of~$G$ with $a(y_v)$, if
$y_v\in\Pi_1\cap Y$, or with $b_{i_x,v}(y_v)$, otherwise. It is easy to see that, analogously to the end of 
the proof of the lower bound for randomized read-once BPs, this gives a protocol for~$\IND_{\ell}$ that
has the same error probability as~$G$. As proved by Klauck~\cite{Kla00}, the complexity of
quantum one-way communication protocols for~$\IND_{\ell}$ with bounded error is lower bounded 
by $\Omega(\ell)$, which together with the facts that only $O(\log p) = O(\log n)$ bits are required
to communicate~$i_x$ and that $\ell \ge n - o(n)$ implies $|G| = 2^{\Omega(n)}$, as claimed.
\end{proof}

\section{The Lower Bound for Set-Disjointness (Theorem~\ref{the:disj})}\label{sec:disj}

In this section, we prove that quantum BPs reading each variable exactly once and computing~$\DISJ_n$ 
with two-sided error bounded by a small positive 
constant require size~$2^{\Omega(n)}$. 
We first present definitions and tools from 
information theory in the next subsection. We then introduce quantum multi-partition protocols 
(Subsection~\ref{sec:multipart}) and prove a lower bound on the information cost of 
such protocols for the AND of just two bits (Subsection~\ref{sec:and}). This is 
used as a building block for the proof of the desired main result in the last subsection.

\vskip0pt plus\baselineskip
\subsection{Information Theory}\label{sec:info_theory}

We assume that the reader is familiar with classical and von Neumann entropy and
refer to~\cite{Nie00} for an introduction. We briefly review some important definitions.

Let $X$ be a classical random variable taking values in a finite set~$R$
and for each $x\in R$ let $\rho(x)$ be a quantum state over a fixed Hilbert space.
Then the state $\rho(X) = {\sum_{x\in R} \Pr\{X=x\}\cdot\rho(x)}$ is 
called \emph{quantum encoding of~$X$ by $(\rho(x))_{x\in R}$}. 
For the special case where $\rho(x) = \k{x}\b{x}$ for each $x\in R$
and $(\k{x})_{x\in R}$ is an ON-basis, we just write~$X$ instead of~$\rho(X)$.
For an additional random variable~$Y$ and a value~$y$ in the range of~$Y$, let
$\rho(X\,|\, Y=y) = {\sum_{x\in R} \Pr\{X=x\,|\, Y=y\}\cdot\rho(x)}$.

For a quantum state~$\rho$, $S(\rho)$ denotes the \emph{von Neumann entropy}
of~$\rho$. For a joint system $(A,B,C)$ with subsystems $A,B,C$, 
define $S(A\,|\,B) = S(A,B) - S(B)$ (\emph{conditional entropy}),
$I(A\,{:}\,B) = S(A)+S(B)-S(A,B)$ (\emph{mutual information between $A$ and~$B$}), 
and $I(A\,{:}\,B\,|\,C) = S(A\,|\,C)+S(B\,|\,C)-S(A,B\,|\,C)$ 
(\emph{conditional mutual information}). For classical random variables $X$, $Y$, and $Z$, a value~$z$ in the
range of~$Z$, and quantum encodings $\rho(X),\sigma(Y)$ of~$X$ and~$Y$, resp., we use the notational 
shortcut $I(\rho(X)\,{:}\,\sigma(Y)\,|\,Z=z) = I(\rho(X\,|\,Z=z)\,{:}\,\sigma(Y\,|\,Z=z))$.
We list the following standard facts for easier reference 
(see, e.\,g., \cite{Nie00}, Sections~11.3--11.4).

\goodbreak
\begin{fact}\label{fact:info}\item[]\vspace*{-2pt}\sloppy\baselineskip=1.05\baselineskip
\begin{shortindent}{(iii)}\newcounter{facti}\renewcommand{\thefacti}{\roman{facti}}\setcounter{facti}{0}
\item[(i)]\refstepcounter{facti}\label{facti:pure_subsys}
  Let $\rho^{AB}$ be a pure state of the joint system $(A,B)$ and let
  $\rho^A,\rho^B$ be the corresponding reduced states of the subsystems $A$ and~$B$, resp.
  Then $S\bigl(\rho^{A}\bigr) = S\bigl(\rho^{B}\bigr)$.
\item[(ii)]\refstepcounter{facti}\label{facti:orth}
  Let $\rho(X) = \sum_{x\in R} \Pr\{X=x\}\cdot\rho(x)$ be a quantum encoding of a classical random variable~$X$
  taking values in the finite set~$R$. Suppose that the states $\rho(x)$, $x\in R$,
  have support on orthogonal subspaces. Then 
  $S(\rho(X)) = H(X) + \sum_{x\in R} \Pr\{X=x\}\cdot S(\rho(x))$,
  where $H(X)$ is the classical entropy of~$X$.
\item[(iii)]\refstepcounter{facti}\label{facti:concavity_ent}
  Let $\rho(X)$ be a quantum encoding of a classical random variable~$X$
  taking values in a finite set~$R$. Then 
  $S(\rho(X)) \ge \sum_{x\in R} \Pr\{X=x\}\cdot S(\rho(x))$
  (concavity of the entropy).
\item[(iv)]\refstepcounter{facti}\label{facti:convex_ent}
  Let $X,Y$ be classical random variables with finite range, let $R$ be the range of~$Y$,
  and let $\rho(X)$ be a quantum encoding of~$X$. Consider a bipartite system with state
  $(\rho(X),Y) = {\sum_{y\in R} \Pr\{Y=y\}\cdot \rho(X\,|\,Y=y)\otimes\k{y}\b{y}}$. Then
  $S(\rho(X)\,|\,Y) = S(\rho(X),Y) - S(Y) = \sum_{y\in R} \Pr\{Y=y\}\cdot S(\rho(X\,|\,Y=y))$.
\item[(v)]\refstepcounter{facti}\label{facti:convex_info}
  Let $X,Y$ be classical random variables with finite range, let $R$ be the range of~$Y$,
  and let $\rho(X),\sigma(X)$ be quantum encodings of~$X$. Consider a tripartite system with 
  state $(\rho(X),\sigma(X),Y) = \sum_{y\in R} \Pr\{Y=y\}\cdot \rho(X\,|\,Y=y)\otimes\sigma(X\,|\,Y=y)\otimes\k{y}\b{y}$. 
  Then $I(\rho(X)\,{:}\,\sigma(X)\,|\,Y) = \sum_{y\in R} \Pr\{Y=y\}\cdot I(\rho(X\,|\,Y=y)\,{:}\,\sigma(X\,|\,Y=y))$.
\item[(vi)]\refstepcounter{facti}\label{facti:mono_mut_info}
  $I(A:B) \le I(A:BC)$ (monotonicity of mutual information).
\item[(vii)]\refstepcounter{facti}\label{facti:concavity_info}
  Let $X=(X_1,\ldots,X_n)$, where $X_1,\ldots,X_n$ are independent 
  classical random variables. Then for any quantum encoding~$\rho(X)$ of~$X$, 
  $I(\rho(X) \,{:}\, X_1,\ldots,X_n) \ge {\sum_{i=1}^n I(\rho(X) \,{:}\, X_i)}$
  (superadditivity of mutual information).
\end{shortindent}
\end{fact}

We observe the following additional property that follows
from the definitions and the fact that the von Neumann entropy 
of pure states is zero.

\begin{fact}\label{fact:info_ent}
Let $\rho(X)$ be a quantum encoding of a classical random variable~$X$ and 
suppose that for each value~$x$ that $X$ can attain, $\rho(x)$ is a pure state.
Then $I(\rho(X)\,{:}\,X) = S(\rho(X))$.
\end{fact}

Furthermore, we work with standard measures for the
distance of quantum states. Let $\rho,\sigma$ be quantum states over
the same Hilbert space. The \emph{trace norm} of $\rho$ is defined
as $\|\rho\|_{\rm t} = \tr|\rho| = \tr\sqrt{\rho^\dag\rho}$ and the
\emph{trace distance} of~$\rho$ and~$\sigma$ as $\|\rho-\sigma\|_{\rm t}$.
The \emph{fidelity} of $\rho$ and~$\sigma$ is 
defined as $F(\rho,\sigma) = \tr\sqrt{\sqrt{\rho}\sigma\sqrt{\rho}}$.
Note that for pure states~$\k{\psi_1}$ and~$\k{\psi_2}$,
$F\bigl(\k{\psi_1}\b{\psi_1},\k{\psi_2}\b{\psi_2}\bigr) = |\bk{\psi_1}{\psi_2}|$.
We will also use the following facts (see, e.\,g., \cite{Nie00}, Section~9.2).

\goodbreak
\begin{fact}\label{fact:dist}\item[]\vspace*{-2pt}\sloppy\hbadness=3000
\begin{shortindent}{(ii)}\newcounter{factii}\renewcommand{\thefactii}{\roman{factii}}\setcounter{factii}{0}
\item[(i)]\refstepcounter{factii}\label{factii:fid_trace}
Let $\k{\psi_1}$, $\k{\psi_2}$ denote pure quantum states. Then 
$
  \bigl\|\k{\psi_1}\b{\psi_1} - \k{\psi_2}\b{\psi_2}\bigr\|_{\rm t}^2
  =  {4\bigl(1 - F\bigl(\k{\psi_1}\b{\psi_1},\k{\psi_2}\b{\psi_2}\bigr)^2\bigr)}.
$
\item[(ii)]\refstepcounter{factii}\label{factii:measure}
Let $\rho_0,\rho_1$ be quantum states and suppose that there is a POV measurement
with boolean results that yields the result~$b\in\{0,1\}$ on state~$\rho_b$
with probability at least~$1-\epsilon$. 
Then $F(\rho_0,\rho_1) \le 2\sqrt{\epsilon(1-\epsilon)}$.
\end{shortindent}
\end{fact}

Further, we note the following ``weak inverse triangle inequality'' for
the inner product of real unit vectors.

\begin{proposition}\label{prop:weak_triangle}\sloppy\hbadness=10000
Let $\k{u},\k{v},\k{w}$ be real unit vectors. Then 
$\bk{u}{w} \ge {2\bigl(\bk{u}{v} + \bk{v}{w}\bigr) - 3}$.
\end{proposition}

\begin{proof}
This follows from 
\[
  \bigl\| \k{u}-\k{w} \bigr\|_2^2 \ \le\  \Bigl(\bigl\|\k{u}-\k{v}\bigr\|_2 + \bigl\|\k{v}-\k{w}\bigr\|_2\Bigr)^2
  \ \le\  2\Bigl(\bigl\|\k{u}-\k{v}\bigr\|_2^2 + \bigl\|\k{v}-\k{w}\bigr\|_2^2\Bigr)
\]
on the one hand and
\[
  \bigl\| \k{u}-\k{w} \bigr\|_2^2 \ =\  2(1-\bk{u}{w})
\]
and similarly for $\bigl\| \k{u}-\k{v} \bigr\|_2^2$, $\bigl\| \k{v}-\k{w} \bigr\|_2^2$ on the other.
\end{proof}

Finally, we need one of the main technical tools from~\cite{Kla01b,Kla04b}
used also in~\cite{Jai03a,Jai03b}. The strong version cited below
has independently been derived in \cite{Kla04b,Jai03b}.

\begin{lemma}[Local transition lemma~\cite{Kla04b,Jai03b}]\label{lem:local}\sloppy
Let $X$ describe a classical uniformly random bit.
Let $\rho_0,\rho_1$ be quantum states over some finite dimensional Hilbert space~$\H$.
Let $\rho(X) = {(\rho_0 + \rho_1)/2}$.
Let $\k{\psi_0}$, $\k{\psi_1}$ be purifications of $\rho_0$ and $\rho_1$, resp., in $\H\otimes\K$,
where $\K$ is a Hilbert space of dimension at least the dimension of~$\H$. 
Then there is a unitary transformation~$U$ on~$\K$
such that for $\k{\psi_0'} = (I\otimes U)\k{\psi_0}$, where $I$ is the identity on~$\H$,
$\bigl\| \k{\psi_1}\b{\psi_1} - \k{\psi_0'}\b{\psi_0'} \bigr\|_{\rm t} \le 2\sqrt{2\,I(\rho(X):X)}$.
\end{lemma}

\subsection{Quantum Multi-Partition Communication Protocols}\label{sec:multipart}

We consider the following simple quantum variant of communication protocols that 
may have more than one input partition. We use quantum one-way communication protocols
with a single input partition as defined, e.\,g., in~\cite{Kre95a}, as building blocks.

\begin{definition}
A \emph{quantum $k$-partition (one-way) communication protocol~$P$} with
respect to nontrivial partitions $\Pi_1,\ldots,\Pi_k$ of the set of input
variables consists of a collection of one-way quantum protocols $P_1,\ldots,P_k$ 
with respect to $\Pi_1,\ldots,\Pi_k$, resp., and
numbers $\alpha_1,\ldots,\alpha_k\in\C$ such that
${|\alpha_1|^2+\cdots+|\alpha_k|^2 = 1}$. Call $\alpha_1,\ldots,\alpha_k$ \emph{initial amplitudes}
of their respective subprotocols.
For $i=1,\ldots,k$ let $\H_i = \H_{i,A}\otimes\H_{i,C}\otimes\H_{i,B}$ be 
the state space of~$P_i$. We require that~$\H_i$ and~$\H_j$ are orthogonal for $i\neq j$.
Let~$\H = \H_1\oplus\cdots\oplus\H_k$ be the global state space of the whole protocol.

The Hilbert space~$\H_i$ describes
the state of a register of qubits on which the subprotocol~$P_i$ works.
For an input $z = (x,y)$ partitioned into $x,y$ according to $\Pi_i$, 
the initial state of the register is $\k{s_i(z)} \ =\ \k{x}_{\H_{i,A}}\k{00\ldots 0}_{\H_{i,C}}\k{y}_{\H_{i,B}}$,
where the three parts of the register belong to the subspaces as indicated. 
The qubits belonging to $\H_{i,A}$ and $\H_{i,B}$, resp., are called the \emph{input registers}
of the players Alice and Bob, resp., and those belonging to $\H_{i,C}$ \emph{work register}.
The computation of~$P_i$ is carried out as usual for quantum one-way protocols.
Let~$U_i$ be the unitary transformation on~$\H_i$ realized by protocol~$P_i$.

The global initial state of~$P$ is $\sum_{i=1}^k \alpha_i \k{s_i(z)}$
and the global final state is $\sum_{i=1}^k \alpha_i U_i\k{s_i(z)}$.
Define $P_1(z),\ldots,P_k(z)$ and $P(z)$, the \emph{result states} of the respective protocols, 
as the states obtained from the respective final states by a partial trace over
the qubits in the input registers of the players. 
The \emph{output random variable of~$P_i$} with values in $\{0,1\}$
is defined as the result obtained by a POV measurement of a designated 
output qubit in~$P_i(z)$ owned by Bob. Let $M_{i,0},M_{i,1}$ be the linear operators
with $M_{i,0}^\dag M_{i,0} + M_{i,1}^\dag M_{i,1} = I$ (the identity on~$\H_i$) that describe
this measurement. Then the \emph{output random variable of~$P$} is the result of the
POV measurement described by the operators $\sum_{i=1}^k M_{i,0}$, $\sum_{i=1}^k M_{i,1}$.
This allows to define the computation of boolean functions with different kinds of error as usual.
A \emph{quantum multi-partition communication protocol} is a
quantum $k$-partition communication protocol for some~$k$.
\end{definition}

{\samepage
\begin{remarks*}\item[]\vspace*{-2pt}
\begin{bulletlist}
\item 
  We do not define the communication complexity of quantum multi-partition 
  protocols here (which can be done in a straightforward way),
  since we measure the complexity using an appropriately defined notion of
  information cost (see the next subsection). 
\item
  Opposed to more generous models of quantum one-way protocols, 
  the subprotocols of our quantum multi-partition protocols are defined such 
  that the players do not obtain any additional, entangled qubits (EPR pairs) as part of the initial 
  state of the protocol. 
\item 
  The above definition can easily be
  generalized by allowing general quantum operations for initialization, more than one round,
  or entanglement between the players. We do not need this kind of generality for our 
  later application, though.
\item 
  Due to the orthogonality of the subspaces of the subprotocols, for each input~$z$, $P(z) = {P_1(z) + \cdots + P_k(z)}$.
  For the same reason, the measurement operators for the output random variable of~$P$ defined above indeed give
  a POV measurement. Finally, let $\O_1,\ldots,\O_k$ and~$\O$ denote the output random variables
  of $P_1,\ldots,P_k$ and~$P$, resp. Then, for $r\in\{0,1\}$, ${\Pr\{ \O = r\}} = {\sum_{i=1}^k |\alpha_i|^2 \Pr\{ \O_i = r \}}$.
\item 
  The initial amplitudes of a quantum multi-partition protocol may be 
  assumed to be real and positive by pushing phase factors into the initial states of the
  subprotocols.
\end{bulletlist}
\end{remarks*}

}

Our goal is to measure the mutual information between the result state
of a protocol and a random input by the simple formula in Fact~\ref{fact:info_ent}.
Hence, it is important that the result state of the considered protocol is pure 
for a given input. At the first glance, this no longer seems to work if we want to
run the protocols on random inputs and want to allow them to use random coins.
The problem is overcome by using input conventions and a simple extension of
the model as described in~\cite{Kla01b,Jai03a}. First, we consider only protocols 
that are safe in the following sense. 

\goodbreak
{\samepage
\begin{definition}[Safe protocols]
A communication protocol is called \emph{safe} if both players may access their 
input registers only once at the beginning to make copies of their 
inputs into the work register. They are not allowed to access the input 
registers for working, communicating, or measuring afterwards. 
\end{definition}

}

It is obvious that requiring protocols to be save does not change their
computational power if we restrict ourselves to classical inputs  as usual.
The convention prevents protocols from entangling their work qubits with the input registers
during the computation, which could lead to the production of extra entropy besides that contained 
in the inputs by the trace-out operation at the end of the computation.

Furthermore, we want to run protocols on random inputs and allow the protocols to use
public random coins, but only want to work with unitary transformations, even for the 
preparation of the initial state. By modifying the model as follows, this is possible.

\begin{definition}[Protocols with random inputs and public random coins]\label{def:mp_ext}
There is an additional \emph{(public) random coin register} whose number of qubits may depend on the length
of the input of Alice and Bob. Furthermore, the input registers of Alice and Bob and the random coin
register are each augmented by a \emph{secret register} of the same size
that are each only initialized once at the beginning and never accessed afterwards.

The protocol is run for random inputs of the two players described by 
random variables $X$ and~$Y$ and random coins described by the random variable~$Z$ as follows. 
At the beginning, Alice prepares the states ${\sum_x \Pr\{X=x\}^{1/2}\k{x}\k{x}}$
and ${\sum_{z} \Pr\{Z=z\}^{1/2}\k{z}\k{z}}$ in the two joint registers formed
by her input register together with its secret register and by the public random coin
register and its secret register (where, e.\,g., the first part 
of each state belongs to the regular register and the second part to the secret one).
Analogously, Bob prepares the state ${\sum_y \Pr\{Y=y\}^{1/2}\k{y}\k{y}}$ in his input register
and the corresponding secret register. 
The \emph{result state} of the protocol is obtained by taking the final 
computational state and tracing out the input registers of both players,
the random coin register, and all secret registers. This is a mixed state which is 
equal to what we would have obtained had we started the protocol on random assignments to the input
registers and the random coin register as described by $X,Y$ and~$Z$, resp., in the first place.
The \emph{output random variable} of such a protocol is the result of a POV measurement of a qubit
owned by Bob at the end, excluding the bits of the random coin register.
\end{definition}

Although the result state according to the extended definition above also depends on~$Z$, 
we stick to the notation $P(X,Y)$ for this state for convenience. 
We summarize the properties of the modified protocols 
that are crucial for the following proofs.

\begin{fact} For a fixed (non-random) assignment to the input registers and the 
random coin register, the result state of a quantum multi-partition protocol as 
described in Definition~\ref{def:mp_ext} is pure. 
Furthermore, for registers initialized with pure states describing random inputs and random 
coins according to the convention in the definition, the computational state at the end
of the protocol before tracing out the input registers, the random coin register,
and the secret registers is also pure.
\end{fact}

\subsection{Information Cost of Quantum Multi-Partition Protocols for AND}\label{sec:and}

Here we prove that the information cost of a quantum multi-partition protocol computing the 
AND of two bits is  lower bounded by a positive constant. For measuring the information cost, 
we adapt the approach of Bar-Yossef, Jayram, Kumar, and Sivakumar~\cite{Yos04a} for 
classical randomized communication protocols and use the information that the result 
state of a protocol provides on the inputs (the result state replacing the classical 
transcript), rather than the weighted sum of the information in individual messages as in
the paper of Jain, Radhakrishnan, and Sen~\cite{Jai03a}. This makes sense also in the quantum 
case since we do not use entanglement and have only a
single round of communication. 

\begin{definition}\item[]\vspace*{-2pt}
\begin{bulletlist}
\item Let $P$ be a quantum $k$-partition protocol. Let $D$ be any random variable and let $Z$ be 
a random variable describing an input for~$P$. Then the \emph{information cost of~$P$ with respect to~$Z$ and
conditioned on~$D$}, denoted by $\IC(P;Z\,|\,D)$, is defined as ${I(P(Z)\,{:}\,Z\,|\,D)}$, where $P(Z)$ is the 
result state of~$P$.
\item \samepage
For a function~$f$, any random variable~$D$, and a random variable~$Z$ describing 
an input for~$f$, the \emph{$\epsilon$-error information cost of quantum $k$-partition
protocols for~$f$ on~$Z$ conditioned on~$D$}, $\IC_{k,\epsilon}(f;Z\,|\,D)$, is defined as the infimum 
of the information cost over all quantum $k$-partition protocols
computing~$f$ with error at most~$\epsilon$. 
Furthermore, let $\IC_{\epsilon}(f;Z\,|\,D) = \min_{k\in\IN} \IC_{k,\epsilon}(f;Z\,|\,D)$ denote
the information cost of quantum multi-partition protocols for~$f$ with error at most~$\epsilon$.
\end{bulletlist}
\end{definition}

To explain some of the difficulties that arise if we want to extend the 
result of Jain, Radha\-krish\-nan, and Sen~\cite{Jai03a} for protocols with a single partition
computing~$\AND$ to multi-partition protocols, we consider the situation for the
$\XOR$ of two bits~$z_1,z_2$.
We choose the following input distribution as defined in~\cite{Yos04a,Jai03a}:
Let $D\in\{1,2\}$ with ${\Pr\{ D = 1\}} = {\Pr\{ D = 2 \}} = 1/2$.
Let $Z=(Z_1,Z_2)$, where for $i=1,2$, $\Pr\{ Z_i = 0 \,|\, D = i \} = \Pr\{ Z_i = 1 \,|\, D = i \} = 1/2$ and 
$\Pr\{ Z_{3-i} = 0 \,|\, D = i \} = 1$. 

\begin{proposition} 
There is an error-free quantum 2-partition protocol for~$\XOR$ on the random input~$Z$ conditioned
on~$D$ where the subprotocols do not communicate at all 
and where each subprotocol has zero information cost with respect to~$Z$ and conditioned on~$D$. 
\end{proposition}

\begin{proof}
By an application of the Deutsch-Jozsa algorithm. We define a 2-partition 
protocol~$P$ according to the partitions $(\{z_1\},\{z_2\})$ and $(\{z_2\},\{z_1\})$. 
The protocol uses two qubits as work space and 
subprotocols $P_1,P_2$ both weighted by the amplitude $1/\sqrt{2}$ (it uses no random coins).
The first work qubit is used for computing, the second one only to implement a phase oracle
as usual. In subprotocol~$P_i$, $i=1,2$, the first work qubit is initialized with $\k{i-1}$.
The only player to act in~$P_i$ is Bob. The only thing he does is multiplying the
phase of the first work qubit by $(-1)^{z_i}$. Then by a measurement of the first
work qubit in the global result state in the Hadamard basis, the value $\XOR(z_1,z_2)$ can 
be retrieved with error probability~$0$. It is easy to check that the
mutual information between the result state~$P_i(Z)$ of subprotocol~$P_i$
and~$Z$ is zero, since Bob only encodes his input in the phase of the work qubit.
\end{proof}

On the other hand, by examining the proof of~\cite{Jai03a} for the $\AND$ of two bits, 
it can be shown that for each quantum 1-partition protocol computing $\XOR$ 
with a bounded number of rounds and with bounded two-sided error, 
the communication complexity as well as the information cost in either the definition
of~\cite{Jai03a} or the definition used here is
lower bounded by a positive constant. In fact, the proof in~\cite{Jai03a} 
only exploits the fact that a protocol for $\AND$ has to be able to distinguish
the inputs $01$ and $10$ from $11$ with high probability and thus works in
the same way for $\XOR$. The example of $\XOR$ and the above proposition show
that a lower bound on the information cost or communication complexity for a single 
partition does not simply carry over to a lower bound for multiple partitions in an 
obvious way. 

As a preparation of the proof of our result for the $\AND$ function,
we state the following concavity property of the information cost
of multi-partition protocols. 

{\samepage
\begin{lemma}\label{lem:decomp_info}\sloppy\hbadness=6000
Let $P$ be a quantum $k$-partition communication protocol with
subprotocols~$P_1,\ldots,P_k$ and initial amplitudes $\alpha_1,\ldots,\alpha_k\in\C$,
where ${|\alpha_1|^2+\cdots+|\alpha_k|^2 = 1}$. Let~$D$ be any random variable
and let $Z$ be a random variable describing a random input for~$P$. Then
$\IC(P;Z\,|\,D) \ge \sum_{i=1}^k |\alpha_i|^2 \IC(P_i;Z\,|\,D)$.
\end{lemma}

}

\begin{proof}
We regard the public random coins of~$P$ 
as part of Alice's input for this proof. 
Then by the definition of the protocols, the result state $P(z)$ for a fixed input~$z$
(which in fact fixes the regular inputs of Alice and Bob, the random coins, and the values
for the secret registers) is a pure state.
Notice, however, that this does \emph{not} mean that the proof only works for pure 
result states. When running~$P$ on the random input~$Z$ and randomly chosen random coins
according to the conventions, the trace-out of the input registers and the corresponding 
secret registers still yields a mixed result state~$P(Z)$.

By definition of the information cost, the statement in the claim is equivalent to
\[
  I(P(Z)\,{:}\,Z\,|\,D) \ \ge\  \sum_{i=1}^k |\alpha_i|^2 I(P_i(Z)\,{:}\,Z\,|\,D).
\]
According to Fact~1(\ref{facti:convex_info}), it suffices to prove this without
the condition on~$D$. Using that the result states of $P_1,\ldots,P_k$ and $P$ are 
pure states for a fixed input and Fact~\ref{fact:info_ent}, it
further suffices to prove that
\[
  S(P(Z)) \ \ge\  \sum_{i=1}^k |\alpha_i|^2 S(P_i(Z)).
\]
For notational convenience, let $p_z = \Pr\{ Z = z\}$ for any input~$z$.
For $i=1,\ldots,k$ let $\k{P_i(z)}$ denote the vector belonging to the pure result state $P_i(z)$
of the subprotocol~$P_i$. Purifying the global result state~$P(Z)$ of~$P$, we obtain
\[
  \k{\psi}
  \ =\  \sum_{i=1}^k \alpha_i \sum_z \sqrt{p_z}\k{P_i(z)}\otimes\k{z}.
\]
Let $\rho = \k{\psi}\b{\psi}$ and let $\rho^A$ and $\rho^B$ be the reduced states 
obtained from $\rho$ by a partial trace over the second and first part, resp.,
of the state space. Using Fact~\ref{fact:info}(\ref{facti:pure_subsys}), we get
\[
  S(P(Z)) \ =\  S\bigl(\rho^A\bigr) \ =\  S\bigl(\rho^B\bigr).
\]
Hence, we investigate $\rho^B$. We have:
\begin{align*}
  \rho^B
  &\ =\  \tr_A\Bigl(\sum_{i,j} \alpha_i\alpha_j^*\sum_{z,z'} \sqrt{p_z}\sqrt{p_{z'}} \k{P_i(z)}\b{P_j(z')}\otimes\k{z}\b{z'}\Bigr)\\
  &\ =\  \sum_{z,z'} \sqrt{p_z}\sqrt{p_{z'}}\k{z}\b{z'}
         \cdot\tr\Bigl(\sum_{i,j} \alpha_i\alpha_j^* \k{P_i(z)}\b{P_j(z')}\Bigr)\\
  &\ =\  \sum_{z,z'} \sqrt{p_z}\sqrt{p_{z'}}\k{z}\b{z'}
         \cdot\sum_i |\alpha_i|^2 \bk{P_i(z')}{P_i(z)}.
\end{align*}
The last row follows from the fact that the state spaces of different subprotocols are
mutually orthogonal. We write the result as
\begin{align*} 
  \rho^B \ =\  \sum_{i=1}^k |\alpha_i|^2 \rho_i\quad\text{with}\quad
  \rho_i \ =\  \sum_{z,z'} \sqrt{p_z}\sqrt{p_{z'}}\bk{P_i(z')}{P_i(z)}\k{z}\b{z'},\;i=1,\ldots,k.
\end{align*}
Define
\[
  \k{\psi_i} \ =\  \sum_{z} \sqrt{p_z} \k{P_i(z)}\otimes\k{z}.
\] 
Then $\rho_i$ is obtained from $\k{\psi_i}\b{\psi_i}$
by tracing over the first part of the state and tracing over
the second yields $P_i(Z)$. Hence, for each~$i$,
$S(\rho_i) \ =\  S(P_i(Z))$, which together with the 
concavity of the entropy (Fact~\ref{fact:info}(\ref{facti:concavity_ent})) proves the claim.
\end{proof}

Next we observe that quantum multi-partition protocols with only two different partitions
can be simplified to quantum $2$-partition protocols. This is obviously applicable to any
quantum multi-partition protocol for a function on just two variables like $\AND$.

\begin{proposition}\label{prop:multi_to_two_part}
Each quantum multi-partition protocol~$P$ with respect to partitions from the set~$\{ \Pi_1,\Pi_2 \}$
can be turned into a quantum $2$-partition protocol~$P'$ with respect to $\Pi_1$ and~$\Pi_2$ 
that has initial amplitudes $\sqrt{q_1},\sqrt{q_2}$ with  $q_1,q_2\ge 0$ and $q_1+q_2 = 1$ and that
for each input has the same result state as~$P$.
\end{proposition}

\begin{proof}
Let $P$ be a quantum $(k_1+k_2)$-partition protocol for $f$ with partitions $\Pi_{1,j} = \Pi_1$ for $j=1,\ldots,k_1$ 
and $\Pi_{2,j} = \Pi_2$ for $j=1,\ldots,k_2$. For $i=1,2$
let $\alpha_{i,1},\ldots,\alpha_{i,k_i}$ be the initial amplitudes of these
partitions and let $q_i = \sum_{j=1}^{k_i} |\alpha_{i,j}|^2$. 

Define a quantum $2$-partition protocol~$P'$ with initial amplitudes $\sqrt{q_1},\sqrt{q_2}$ and 
partitions $\Pi_1$ and $\Pi_2$ as follows. For $i=1,2$ 
and $j=1,\ldots,k_i$ let $\k{s_{i,j}}$ be the initial state of the subprotocol~$P_{i,j}$ with partition $\Pi_{i,j}$ in~$P$. 
Then for $i=1,2$ the initial state of the subprotocol~$P_i'$ of~$P'$ with partition~$\Pi_i$ is defined as 
$\sum_j (\alpha_{i,j}/\sqrt{q_i})\k{s_{i,j}}$, if $q_i\neq 0$,
or as an arbitrary pure state, if $q_i = 0$. In this way, we get a legal pure state
that can be prepared by Alice at the beginning of the computation of the $i$th subprotocol. 
In $P_i'$, the players then simulate the respective subprotocols $P_{i,1},\ldots,P_{i,k_i}$ of~$P$ 
in parallel. By the definitions it is obvious that, for each input, the final computation state of~$P'$ 
agrees with that of~$P$. Hence, the same follows also for the result states.
\end{proof}

Furthermore, we observe that it suffices to work with real amplitudes in the protocols.
For a complex vector space with basis $b_1,\ldots,b_n$, its \emph{realification} is the
real vector space spanned by the basis $b_1,\ldots,b_n,ib_1,\ldots,ib_n$
(using the operations of the complex vector space but allowing only real scalars). 
The \emph{realification of a complex vector} is obtained by replacing 
each of its entries with two entries containing its real and imaginary part, resp. 
To get the \emph{realification of a complex matrix}, replace each of 
its entries~$a$ with a $2\times 2$-block 
$\left(\begin{smallmatrix}b_1&b_2\\b_3&b_4\end{smallmatrix}\right)$ 
where $b_1 = b_4 = \Re(a)$, $b_3 = -b_2 = \Im(a)$.
The \emph{realification of a quantum state} $\rho = \sum_{i=1}^n p_i \k{\psi_i}\b{\psi_i}$,
with ${p_1,\ldots,p_n\ge 0}$, $p_1+\cdots+p_n = 1$, and $\k{\psi_1},\ldots,\k{\psi_n}$ an ON-basis,
is the quantum state $\rho' = \sum_{i=1}^n p_i \k{\psi_i'}\b{\psi_i'}$ where $\k{\psi_i'}$ is 
the realification of $\k{\psi_i}$ for $i\in\{1,\ldots,n\}$. Finally, the \emph{realification of 
a quantum communication protocol} is the protocol resulting from the 
replacement of the initial state as well as of the matrices describing the computation and the 
measurements of the protocol with their realifications. 
It is easy to see (see also~\cite{Kre95a}, Lemma~6) 
that the final state of the resulting protocol is then the realification of the 
original final state. For our purposes, we require the following additional fact.

\begin{fact}\label{fact:ent_real}
The von Neumann entropy of a quantum state agrees with that of its realification. 
In particular, the information theoretical measures introduced at the beginning 
of the section are preserved if all involved states are replaced with their realifications.
\end{fact}

\begin{proof}
Let $\rho$ and $\rho'$ be a quantum state and its realification, resp., as defined above.
Then the realifications of the vectors $\k{\psi_1},\ldots,\k{\psi_n}$ and 
$i\k{\psi_1},\ldots,i\k{\psi_n}$ constitute an ON-basis of eigenvectors of~$\rho'$
with corresponding eigenvalues $p_1,\ldots,p_n$ and the eigenvalue~$0$ with multiplicity~$n$.
Hence, $S(\rho') = S(\rho)$. The second part of the claim is obvious.
\end{proof}

Now we consider the~$\AND$ of two bits~$z_1,z_2$.
We consider the same input distribution for~$\AND$ as described before for~$\XOR$.
Recall that $D\in\{1,2\}$ with $\Pr\{ D = 1\} = {\Pr\{ D = 2 \}} = 1/2$ and that
$Z=(Z_1,Z_2)$, where for $i=1,2$, $\Pr\{ Z_i = 0 \,|\, D = i \} = \Pr\{ Z_i = 1 \,|\, D = i \} = 1/2$ 
and $\Pr\{ Z_{3-i} = 0 \,|\, D = i \} = 1$. 
We are now ready to state and prove the main theorem of this subsection.

\begin{theorem}\label{the:and}\sloppy
Let $\epsilon\ge 0$ and $\delta = 2\sqrt{\epsilon(1-\epsilon)}$
be such that $\delta \le 1/7$ $($or, equivalently, 
${\epsilon \le 1/2 - 2\sqrt{3}/7} \approx 0.005$$)$.
Then $\IC_{\epsilon}(\AND;Z\,|\,D) \ge 1/28 - \delta/4$.
\end{theorem}

The plan for the proof of the theorem is as follows. 
We have to show that if a given protocol~$P$ computes~$\AND$ with small error probability, then $I(P(Z)\,{:}\,Z\,|\,D)$ is large. 
First, we can restrict ourselves to $2$-partition protocols using Proposition~\ref{prop:multi_to_two_part}.
We then apply Lemma~\ref{lem:decomp_info} to lower bound the overall information $I(P(Z)\,{:}\,Z\,|\,D)$ 
by the average of that given by the subprotocols~$P_1,P_2$ of~$P$. Due to the known
results, it is clear that the information provided by an individual, single-partition subprotocol about 
a random input of the considered kind is large if it computes~$\AND$ with small error probability. But this does
not suffice to conclude the proof, as the example of $\XOR$ discussed above shows. The problem
is that, in general, having a protocol $P$ with small overall error probability for each input does not imply
that there is a subprotocol which shares this property. As a way around this problem, we use the fidelity as a measure
for the ability of the protocols to distinguish between the inputs~$00$, $01$, and $10$ on the one
hand and the input~$11$ on the other. Using the properties of the fidelity, we can show that if the
whole protocol can reliably distinguish between these sets of inputs, which it has to if its
error probability is to be small, then the same is true for at least one of the subprotocols.
The local transitition lemma then in turn implies that this subprotocol provides a 
nonnegligible amount of information about a random input as chosen above. 
We now make this more precise.

\begin{proof}[Proof of Theorem~\ref{the:and}]
Due to Proposition~\ref{prop:multi_to_two_part}, we may assume that the given
protocol for~$\AND$ is a 2-partition protocol with respect to the
partitions $\Pi_1 = (\{z_1\},\{z_2\})$ and $\Pi_2 = (\{z_2\},\{z_1\})$. 
Let~$P$ be such a protocol computing $\AND$ with error at most~$\epsilon$.
Let $P_1,P_2$ be the subprotocols of~$P$ that have initial amplitudes 
$\alpha_1 = \sqrt{q_1}, \alpha_2 = \sqrt{q_2}$ with $q_1,q_2\ge 0$ and $q_1+q_2 = 1$.
Furthermore, because of Fact~\ref{fact:ent_real}, we may additionally assume
that~$P$ uses only real numbers in its transition and measurement
matrices as well as in its computational states.

Let~$Z = (Z_1,Z_2)$ be the input random variable for~$P$ as defined before and 
let $Z_{1,j} = Z_j$ and $Z_{2,j} = Z_{3-j}$ for $j=1,2$. We denote the result state 
of~$P$ on~$Z$ by~$P(Z)$.
By Lemma~\ref{lem:decomp_info} and Fact~\ref{fact:info}(\ref{facti:mono_mut_info})
(the latter together with Fact~\ref{fact:info}(\ref{facti:convex_info}) for handling the 
additional condition on~$D$),
\begin{align*}
  I(P(Z)\,{:}\,Z\,|\,D) 
  &\ \ge\  q_1 I(P_1(Z)\,{:}\,Z\,|\,D) + q_2 I(P_2(Z)\,{:}\,Z\,|\,D)\\
  &\ \ge\  q_1 I(P_1(Z)\,{:}\,Z_{1,1}\,|\,D) + q_2 I(P_2(Z)\,{:}\,Z_{2,1}\,|\,D).
\end{align*}
{\sloppy\hbadness=2700
Furthermore, due to the fact that $Z_{i,1}$ conditioned on $D = 3-i$ is the fixed bit~$0$,
$I(P_i(Z)\,{:}\,Z_{i,1}\,|\,D) = (1/2)I(P_i(Z)\,{:}\,Z_{i,1}\,|\,D=i)$.
For $i=1,2$ let $\eta_i = {I(P_i(Z)\,{:}\,Z_{i,1}\,|\,D=i)}$. Altogether, we have shown that
\setcounter{equation}{0}
\begin{equation}\label{eq:and_i}
  I(P(Z)\,{:}\,Z\,|\,D) \ \ge\ \frac{1}{2}(q_1\eta_1+q_2\eta_2).
\end{equation}
Our goal is to lower bound the right hand side in terms of the error probability of the protocol~$P$.

}

We analyze $\eta_1 = I(P_1(Z)\,{:}\,Z_{1,1}\,|\,D=1)$ in detail.
Observe that, conditioned on ${D = 1}$, $P_1(Z) = P_1(Z_{1,1},0)$ and $Z_{1,1}$ is 
a uniformly random bit.
We also run~$P_1$ on the fixed (non-random) input $(b_1,b_2)\in\{0,1\}^2$, which means that,
according to our conventions, the players Alice and Bob prepare states
$\k{b_1}\k{b_1}\bigl(\sum_{z}\sqrt{p_{z}}\k{z}\k{z}\bigr)$ and $\k{b_2}\k{b_2}$,
resp. The first two parts of each state correspond to the regular input register and 
its secret register. The second two parts of Alice's state are the contents of
the public random coin register and its secret register, where
$(p_z)_z$ is the distribution of the values for the random coins. 
Let $\k{s_1(b_1,b_2)}$ be the final computational state of~$P_1$
on input~$(b_1,b_2)\in\{0,1\}^2$, before tracing out any register. 
It is obvious that this is a pure state. 

Let \emph{Alice's extended input register} be the joint register consisting of Alice's 
input register, the public random coin register, and the respective secret registers.
Let $P_1'(b_1,b_2)$ be the state obtained from~$\k{s_1(b_1,b_2)}$ 
by tracing out Alice's extended input register.
In general, the obtained state is mixed due to the random coin component. 
We may regard the states
$\k{s_1(00)}$, $\k{s_1(10)}$ as purifications of the states 
$P_1'(00)$, $P_1'(10)$, resp., where the Hilbert space of Alice's extended input register
serves as the extension space. Since conditioned on $D = 1$, Bob's part of~$Z$ is the 
fixed input~$0$, we have ${I(P_1'(Z)\,{:}\,Z_{1,1}\,|\,D=1)} = I(P_1(Z)\,{:}\,Z_{1,1}\,|\,D=1)$.

Now we apply the local transition lemma (Lemma~\ref{lem:local})
to the states $\rho_0 = P_1'(00)$ and $\rho_1 = P_1'(10)$
and their purifications $\k{s_1(00)}$ and~$\k{s_1(10)}$, resp.
Observe that $P_1'(Z\,|\,D=1) = P_1'(Z_{1,1},0) = (1/2)(P_1'(00)+P_1'(10))$.
Due to the lemma, there is a unitary correction
transformation~$V$ acting nontrivially only on the Hilbert space of
Alice's extended input register such that
\[
  \bigl\|V \k{s_1(00)} - \k{s_1(10)}\bigr\|_{\rm t} 
  \ \le\  2\sqrt{2\,I(P_1'(Z):Z_{1,1}|D=1)} 
  \ =\  2\sqrt{2\eta_1},
\]
where for the sake of readability, pure states are only written as vectors.
Let $U(z)$ be the unitary transformation applied by Bob in the protocol~$P_1$
if his input bit is~$z$. 
Then, by the unitary invariance of the trace norm and the fact that $U(z)$, $z\in\{0,1\}$,
and $V$ commute:
\begin{align*}
  \bigl\|V \k{s_1(01)}  - \k{s_1(11)}\bigr\|_{\rm t}
  &\ =\  
  \bigl\|V U(1)U(0)^\dag\k{s_1(00)} - U(1)U(0)^\dag\k{s_1(10)}\bigr\|_{\rm t}\\
  &\ =\  \bigl\|V \k{s_1(00)} - \k{s_1(10)}\bigr\|_{\rm t} \ \le\  2\sqrt{2\eta_1}.
\end{align*}
As abbreviations, let $\k{s_1'(00)} = V\k{s_1(00)}$ and $\k{s_1'(01)} = V\k{s_1(01)}$.
Observe that 
\begin{align*}
  \bk{s_1'(01)}{s_1(11)} 
  &\ =\  \b{s_1(01)}V^\dag\k{s_1(11)}\\
  &\ =\  \b{s_1(00)}U(0)U(1)^\dag V^\dag\cdot U(1)U(0)^\dag\k{s_1(10)}\\
  &\ =\  \bk{s_1'(00)}{s_1(10)}.
\end{align*}
Using the relationship between fidelity and trace distance for pure states from 
Fact~\ref{fact:dist}(\ref{factii:fid_trace}) and setting $\gamma_1 = 2\eta_1$ as 
an abbreviation, it follows that
\begin{equation}\label{eq:and_ii}
  F\bigl(\k{s_1'(01)},\k{s_1(11)}\bigr)
  \ =\  F\bigl(\k{s_1'(00)},\k{s_1(10)}\bigr)
  \ \ge\  \sqrt{1-2\eta_1}
  \ \ge\  1-\gamma_1.
\end{equation}
We treat the subprotocol~$P_2$ in the same way. Notice that the input of~$P_2$ is also $Z = (Z_1,Z_2)$, 
but now Alice has~$Z_2$ and Bob has~$Z_1$. Conditioned on $D = 2$, $Z_2$ is a random bit and $Z_1 = 0$.
Let $\k{s_2(b_1,b_2)}$ be the final computational state of~$P_2$ on input $(b_1,b_2)\in\{0,1\}^2$
(before tracing out any register). Again, this is a pure state. Let $\k{s_2'(00)}$ and $\k{s_2'(01)}$ 
be the states resulting from the application of a correction transformation according to 
the local transition lemma. Let $\gamma_2 = 2\eta_2$. 
Then, analogously to the above, $\bk{s_2'(10)}{s_2(11)} = \bk{s_2'(00)}{s_2(01)}$ and
\begin{equation}\label{eq:and_iii}
  F\bigl(\k{s_2'(10)},\k{s_2(11)}\bigr) \ =\  F(\k{s_2'(00)},\k{s_2(01)}) \ \ge\  1-\gamma_2.
\end{equation}

{\sloppy
We still have to connect the \emph{local} information about the 
subprotocols that we have just derived to the \emph{global} behavior of the protocol~$P$
in order to exploit the fact that $P$ computes~$\AND$ with small error probability.
For this, we first relate the distances of the states for the
subprotocols to those for the whole protocol. 
Let $\k{s(00)} = \alpha_1\k{s_1'(00)} + \alpha_2\k{s_2'(00)}$,
$\k{s(01)} = \alpha_1\k{s_1'(01)} + \alpha_2\k{s_2(01)}$,
$\k{s(10)} = {\alpha_1\k{s_1(10)} + \alpha_2\k{s_2'(10)}}$, and
$\k{s(11)} = \alpha_1\k{s_1(11)} + \alpha_2\k{s_2(11)}$.
Then, using that $P_1$ and $P_2$ work on orthogonal subspaces
and the definitions $|\alpha_1|^2 = q_1$, $|\alpha_2|^2 = q_2$, we get
\begin{align}
  \bk{s(01)}{s(11)}
    &\ =\  q_1\bk{s_1'(01)}{s_1(11)} + q_2\bk{s_2(01)}{s_2(11)},\label{eq:and_iv}\\
  \bk{s(10)}{s(11)}
    &\ =\  q_1\bk{s_1(10)}{s_1(11)} + q_2\bk{s_2'(10)}{s_2(11)},\;\text{and}\label{eq:and_v}\\
  \bk{s(00)}{s(11)}
    &\ =\  q_1\bk{s_1'(00)}{s_1(11)} + q_2\bk{s_2'(00)}{s_2(11)}.\label{eq:and_vi}
\end{align}

}

{\sloppy
Second, we relate the absolute value of the left hand sides of
equations \mbox{(\ref{eq:and_iv})--(\ref{eq:and_vi})}, i.\,e., the
fidelity of the respective pairs of states, to the error probability of the protocol~$P$.
Let $\rho$ be a quantum state over the Hilbert space of the computational states of~$P$ and 
let $\tr_{\,\rm non-work}(\rho)$ denote the state obtained from~$\rho$ 
by tracing out the input registers of both players, 
the random coin register, and the respective secret registers. Since the correction 
transformations only work on Alice's extended input register or on Bob's extended input register,
their effect disappears after applying $\tr_{\,\rm non-work}$.
Hence, for $(b_1,b_2)\in\{0,1\}^2$,
\[
  \tr_{\,\rm non-work}\bigl(\k{s(b_1,b_2)}\b{s(b_1,b_2)}\bigr) \ =\  P(b_1,b_2).
\]
Furthermore, the qubit measured to get the output of the
protocol does not belong to the bits traced out in this way. 
We can thus apply Fact~\ref{fact:dist}(\ref{factii:measure}) to
get $F\bigl(\k{s(01)},\k{s(11)}\bigr) \le \delta$ for $\delta = 2\sqrt{\epsilon(1-\epsilon)}$,
where $\epsilon$ is the error probability of~$P$.
Analogously, $F\bigl(\k{s(10)},\k{s(11)}\bigr) \le \delta$ and 
$F\bigl(\k{s(00)},\k{s(11)}\bigr) \le \delta$.

}

Third, we prove the desired relationship between error 
probability and information about the inputs stored in the subprotocols. 
Here it is crucial that the considered state vectors only have real components.

\begin{claim*}
Let $\tau > 0$ and suppose that
$\delta \le 1 - {(q_1\gamma_1+q_2\gamma_2)} - {(3/2)(1/2+\tau)}$.
\begin{shortindent}{2.\ }\itemsep2pt
\item[1.] Suppose that there is an $i\in\{1,2\}$ such that $q_i\ge 1/2+\tau$. 
  Then $\delta \ge {2\tau-(1/2+\tau)\gamma_i}$.
\item[2.] Let $q_1,q_2\in[1/2-\tau,1/2+\tau]$.\vspace*{2pt}
  \begin{shortindent}{2.2.}\itemsep2pt
  \item[2.1.] Suppose that both terms on the right hand side of
    equation~(\ref{eq:and_iv}) or~(\ref{eq:and_v}), resp., have the same sign.
    Then $\delta \ge (1/2-\tau)(1-\gamma_1)$ or $\delta \ge {(1/2-\tau)(1-\gamma_2)}$, resp.
  \item[2.2.] Otherwise, $\delta \ge 1/5 - (4/5)(q_1\gamma_1+q_2\gamma_2)$.
  \end{shortindent}
\end{shortindent}
\end{claim*}

\begin{proof}
\emph{Case 1:} W.\,l.\,o.\,g., $q_1 \ge 1/2+\tau$ and thus $q_2\le 1/2-\tau$. 
By equation~(\ref{eq:and_iv})
and the lower bound on the fidelity from~(\ref{eq:and_ii}),
\begin{align*}
  \delta 
  &\ \ge\ F\bigl(\k{s(01)},\k{s(11)}\bigr)
  \ \ge\ q_1|\bk{s_1'(01)}{s_1(11)}| - q_2|\bk{s_2(01)}{s_2(11)}|\\
  &\ \ge\ q_1 F\bigl(\k{s_1'(01)}, \k{s_1(11)}\bigr) - q_2 
   \ \ge\  q_1(1-\gamma_1) - q_2
   \ \ge\ 2\tau - (1/2+\tau)\gamma_1.
\end{align*}

\medskip
\emph{Case 2.1:}
Let, e.\,g., equation~(\ref{eq:and_iv}) have solely nonnegative terms
on its right hand side. Then  
\begin{align*}
  \delta 
  &\ \ge\ F\bigl(\k{s(01)},\k{s(11)}\bigr)
  \ =\ q_1|\bk{s_1'(01)}{s_1(11)}| + q_2|\bk{s_2(01)}{s_2(11)}|\\
  &\ \ge\ q_1(1-\gamma_1) \ \ge\  (1/2-\tau)(1-\gamma_1).
\end{align*}

\medskip
\emph{Case 2.2:}
This is split into two further subcases handling the different possible
signs of the inner products in equations~(\ref{eq:and_iv})
and (\ref{eq:and_v}). 

\emph{Case 2.2.1:}  Suppose first that the first terms on the right hand sides of
equation~(\ref{eq:and_iv}) and~(\ref{eq:and_v}) have the same sign. Then due to
the case distinction, the second terms have the respective opposite sign.
We assume that $\bk{s_1'(01)}{s_1(11)}\ge 0$ and $\bk{s_1(10)}{s_1(11)}\ge 0$
(the other case, $\bk{s_1'(01)}{s_1(11)}\le 0$ and $\bk{s_1(10)}{s_1(11)}\le 0$,
is handled analogously). We claim that then both inner products
on the right hand side of equation~(\ref{eq:and_vi}) are nonnegative and 
of large absolute value.
Using the weak inverse triangle inequality for inner products of real vectors
(Proposition~\ref{prop:weak_triangle}), we get
\begin{equation}\label{eq:and_vii}\tag{$*$}
  \bk{s_1'(00)}{s_1(11)}
  \ \ge\  2\bigl(\bk{s_1'(00)}{s_1(10)} + \bk{s_1(10)}{s_1(11)}\bigr) - 3.
\end{equation}
(If treating the case that $\bk{s_1'(01)}{s_1(11)}\le 0$ and 
$\bk{s_1(10)}{s_1(11)}\le 0$, apply the inverse triangle inequality
to the vectors $\k{s_1'(00)}$, $-\k{s_1(10)}$, and $\k{s_1(11)}$ and
otherwise proceed in the same way as described here.)
Due to the lower bound on the fidelity from~(\ref{eq:and_ii}) and
since $\bk{s_1'(01)}{s_1(11)}\ge 0$,
\[
  \bk{s_1'(00)}{s_1(10)} \ =\  \bk{s_1'(01)}{s_1(11)} \ \ge\  1-\gamma_1.
\]
Furthermore, using that $\bk{s_1(10)}{s_1(11)}\ge 0$, $\bk{s_2'(10)}{s_2(11)} < 0$, and
equation~(\ref{eq:and_v}), we get 
\[
  |\bk{s_1(10)}{s_1(11)}| \ =\  \bk{s_1(10)}{s_1(11)}
  \ \ge\ \frac{1}{q_1}\bigl(q_2|\bk{s_2'(10)}{s_2(11)}| - \delta\bigr),
\]
which together with the lower bound on the fidelity from fact~(\ref{eq:and_iii}) implies
\[
  |\bk{s_1(10)}{s_1(11)}| 
  \ \ge\ \frac{q_2}{q_1}(1-\gamma_2) - \frac{1}{q_1}\delta.
\]
Substituting this into (\ref{eq:and_vii}) yields
\begin{align*}
  \bk{s_1'(00)}{s_1(11)}
  &\ \ge\  2\Bigl(1-\gamma_1 + \frac{q_2}{q_1}(1-\gamma_2) - \frac{1}{q_1}\delta\Bigr) - 3
  \ =\  \frac{2}{q_1}\bigl(1-(q_1\gamma_1+q_2\gamma_2) - \delta\bigr) - 3.
\end{align*}
{\hbadness=2500
Next we apply the inverse triangle inequality to the vectors $\k{s_2'(00)}$, $-\k{s_2(01)}$, and $\k{s_2(11)}$.
Recall that in the considered case, $\bk{s_2(01)}{s_2(11)} < 0$ and $\bk{s_2'(10)}{s_2(11)} = \bk{s_2'(00)}{s_2(01)} < 0$. 
Analogously to the calculations above, we get
\begin{align*}
  \bk{s_2'(00)}{s_2(11)}
  &\ \ge\  2\bigl(\underbrace{-\bk{s_2'(00)}{s_2(01)}}_{\ge\ 1-\gamma_2} +
     \underbrace{\bigl(-\bk{s_2(01)}{s_2(11)}\bigr)}_{\ge\ \frac{q_1}{q_2}(1-\gamma_1)-\frac{1}{q_2}\delta}\bigr) - 3\\
  &\ =\  \frac{2}{q_2}\bigl(1-(q_1\gamma_1+q_2\gamma_2) - \delta\bigr) - 3.
\end{align*}}%
By the derived estimates and the fact that 
\[
  \delta \ \le\  1 - (q_1\gamma_1+q_2\gamma_2) - \frac{3}{2}\Bigl(\frac{1}{2}+\tau\Bigr)
  \ \le\  1 - (q_1\gamma_1+q_2\gamma_2) - \frac{3}{2}\max\{q_1,q_2\}
\]
due to the hypothesis of the claim, the inner products $\bk{s_1'(00)}{s_1(11)}$
and $\bk{s_2'(00)}{s_2(11)}$ are both nonnegative.
Hence, using equation~(\ref{eq:and_vi}) we obtain
\begin{align*}
  \delta 
  &\ \ge\  q_1|\bk{s_1'(00)}{s_1(11)}| + q_2|\bk{s_2'(00)}{s_2(11)}|\\
  &\ \ge\ 4\bigl(1-(q_1\gamma_1+q_2\gamma_2) - \delta\bigr) - 3,\;\;\text{and solving for~$\delta$,}\\
  \delta
  &\ \ge\  \frac{1}{5} - \frac{4}{5}(q_1\gamma_1+q_2\gamma_2).
\end{align*}
This completes the proof for Case~2.2.1.

\medskip
\emph{Case 2.2.2:} In the last remaining case,
the first terms on the right hand side of equation~(\ref{eq:and_iv}) 
and~(\ref{eq:and_v}) have opposite sign. 
W.\,l.\,o.\,g., let $\bk{s_1'(01)}{s_1(11)}\ge 0$
and $\bk{s_1(10)}{s_1(11)}\le 0$. We claim that then both inner products
on the right hand side of equation~(\ref{eq:and_vi}) are nonpositive
and of large absolute value.
Applying the weak inverse triangle inequality for inner products
to the vectors $\k{s_1'(00)}$, $\k{s_1(10)}$, $-\k{s_1(11)}$ and to
the vectors $\k{s_2'(00)}$, $\k{s_2(01)}$, $-\k{s_2(11)}$, resp., yields
\begin{align*}
  -\bk{s_1'(00)}{s_1(11)}
  &\ \ge\  2\bigl(\bk{s_1'(00)}{s_1(10)} + \bigl(-\bk{s_1(10)}{s_1(11)}\bigr)\bigr) - 3\;\;\text{and}\\
  -\bk{s_2'(00)}{s_2(11)}
  &\ \ge\  2\bigl(\bk{s_2'(00)}{s_2(01)} + \bigl(-\bk{s_2(01)}{s_2(11)}\bigr)\bigr) - 3,\;\text{resp.}
\end{align*}
Using arguments analogous to case~2.2.1, the right hand sides of these inequalities can be
lower bounded by nonnegative expressions in $\gamma_1,\gamma_2,\delta$.
The ensuing calculations are also analogous to the above ones, giving the same lower
bound on~$\delta$ in terms of $\gamma_1,\gamma_2$.
\end{proof}

{\sloppy\hbadness=2000
Finally, it only remains to exploit the bounds on $\delta$ in terms of $\gamma_1,\gamma_2$
due to the claim to get bounds on the information cost of the whole protocol. We split
the computation into cases as in the claim. Due to equation~(\ref{eq:and_i})
and taking into account that $\eta_i = \gamma_i/2$ for $i=1,2$, we have
\[
  I(P(Z)\,{:}\,Z|D) \ \ge\  \frac{1}{2}(q_1\eta_1 + q_2\eta_2) \ =\  \frac{1}{4}(q_1\gamma_1 + q_2\gamma_2).
\]
For the following case distinction, we assume that the hypothesis of the claim,
$\delta \le {1  - (q_1\gamma_1+q_2\gamma_2) - (3/2)(1/2+\tau)}$, is satisfied.

}

\medskip
{\sloppy
\emph{Case 1:} Again, we only consider the subcase $q_1 \ge 1/2+\tau$. Due to the
claim, $\delta \ge {2\tau - (1/2+\tau)\gamma_1}$, implying
$\gamma_1 \ge (1/2+\tau)^{-1}(2\tau-\delta)$. Thus, using that $q_1 \ge 1/2+\tau$, 
\begin{equation}\label{eq:and_viii}
  I(P(Z)\,{:}\,Z\,|\,D) 
  \ \ge\  \frac{1}{4}q_1\gamma_1
  \ \ge\  \frac{1}{4}\Bigl(\frac{1}{2}+\tau\Bigr)\Bigl(\frac{1}{2}+\tau\Bigr)^{-1}(2\tau-\delta)
  \ =\  \frac{1}{2}\tau - \frac{1}{4}\delta.
\end{equation}

}

\medskip
\emph{Case 2.1:} W.\,l.\,o.\,g., let $\delta \ge (1/2-\tau)(1-\gamma_1)$ by the claim,
i.\,e., $\gamma_1 \ge 1-(1/2-\tau)^{-1}\delta$. Then, using that $q_1\ge 1/2-\tau$ in this case,
\begin{equation}\label{eq:and_ix}
  I(P(Z)\,{:}\,Z\,|\,D) 
  \ \ge\  \frac{1}{4}q_1\gamma_1
  \ \ge\  \frac{1}{4}\Bigl(\frac{1}{2}-\tau\Bigr)\Bigl(1-\Bigl(\frac{1}{2}-\tau\Bigr)^{-1}\delta\Bigr)
  \ =\  \frac{1}{8} - \frac{1}{4}\tau - \frac{1}{4}\delta.
\end{equation}

\medskip
\emph{Case 2.2:} We have $\delta \ge 1/5 - (4/5)(q_1\gamma_1+q_2\gamma_2)$ by the claim,
i.\,e., $q_1\gamma_1+q_2\gamma_2 \ge 1/4 - (5/4)\delta$. Then
\begin{equation}\label{eq:and_x}
  I(P(Z)\,{:}\,Z\,|\,D) 
  \ \ge\  \frac{1}{4}(q_1\gamma_1+q_2\gamma_2)
  \ \ge\  \frac{1}{4}\Bigl(\frac{1}{4}-\frac{5}{4}\delta\Bigr)
  \ =\  \frac{1}{16} - \frac{5}{16}\delta.
\end{equation}

\medskip
We still have to take the upper bound on~$\delta$ needed for the
application of the claim into account. This requires that
\[
  \delta \ \le\  1  - (q_1\gamma_1+q_2\gamma_2) - \frac{3}{2}\Bigl(\frac{1}{2}+\tau\Bigr).
\]
Since $I(P(Z)\,{:}\,Z\,|\,D) \ge (1/4)(q_1\gamma_1+q_2\gamma_2)$, the above is satisfied if
\[
  I(P(Z)\,{:}\,Z\,|\,D) \ \le\  \frac{1}{16} - \frac{3}{8}\tau - \frac{1}{4}\delta .
\]

Now either the assumption of the lemma is not satisfied and negating the last inequality gives us
the lower bound
\begin{equation}\label{eq:and_xi}
  I(P(Z)\,{:}\,Z\,|\,D) \ \ge\  \frac{1}{16} - \frac{3}{8}\tau - \frac{1}{4}\delta,
\end{equation}
or we get the minimum of (\ref{eq:and_viii})--(\ref{eq:and_x}) as a lower bound.
It remains to fix~$\tau$ such that we get a positive lower bound on the information for
the largest possible~$\delta$.

We choose $\tau = 1/14$ and assume that $\delta \le 1/7$. Then either~(\ref{eq:and_xi}) is satisfied and
thus $I(P(Z)\,{:}\,Z\,|\,D) \ge 1/28 - (1/4)\delta$, or the claim is applicable and
a lower bound is given by the minimum of (\ref{eq:and_viii})--(\ref{eq:and_x}),
\[
  \min\Bigl\{ \frac{1}{28} - \frac{1}{4}\delta, \frac{3}{28} - \frac{1}{4}\delta, \frac{1}{16} - \frac{5}{16}\delta \Bigr\}.
\]
The last term in the minimum is smaller than the first one only if $\delta > 3/7$. Since we have assumed that $\delta \le 1/7$,
we again get the lower bound $I(P(Z)\,{:}\,Z\,|\,D) \ge 1/28 - (1/4)\delta$. 

\sloppy
Altogether, we have shown that for $\delta \le 1/7$,
\[
  I(P(Z)\,{:}\,Z|D) 
  \ \ge\  \frac{1}{28} - \frac{1}{4}\delta.
\]
The claim on the range of the error probabilities in the theorem follows by 
substituting $\delta = 2\sqrt{\epsilon(1-\epsilon)}$ into the bound $\delta < 1/7$.
\end{proof}

\subsection{Application to Quantum Read-Once BPs for the Disjointness Function}

It is convenient here to work with the negation of~$\DISJ_n$, i.\,e., the non-disjointness function
defined by $\ND_n(x,y) = {x_1y_1\lor\cdots\lor x_ny_n}$.
We consider the same input distribution for $\ND_n$ as in~\cite{Yos04a,Jai03a}.
Let $D$ and $Z$ be random variables as defined for $\AND$ in the previous subsection. 
Let $\vec{D} = (D_1,\ldots,D_n)$ and $\vec{Z} = (Z_1,\ldots,Z_n)$ consist
of $n$ independent copies of $D$ and $Z$, resp. For $i=1,\ldots,n$ let $Z_i = (Z_{i,1},Z_{i,2})$.
Observe that, for any value~$\vec{d}$ that~$\vec{D}$ can attain and for each $i=1,\ldots,n$,
the random variables $Z_{i,1},Z_{i,2}$ are independent
when conditioned on $\vec{D} = \vec{d}$. 
Furthermore, if for any~$i\in\{1,\ldots,n\}$ and any $(a,b)\in\{0,1\}^2$ we let $\vec{Z}'$ be the modified input
obtained from~$\vec{Z}$ by replacing $Z_i$ with~$(a,b)$, we observe that, with probability~$1$, $\ND_n(\vec{Z}') = \AND(a,b)$.

Call a quantum read-once BP \emph{regular} if it reads each of its variables at least once,
i.\,e., on each of its paths each variable occurs exactly once. Observe that, in particular, this means that 
such a graph is leveled. For a regular quantum read-once BP~$G$ and an input~$z$ of length~$n$, let $G(z)$ denote 
the final quantum state computed by $G$ on~$z$ after~$n$ computation steps, before the measurement at the sinks 
occurs. Observe that~$G(z)$ is a pure state by the definition of QBPs. 
Using Fact~\ref{fact:info_ent}, we get:

\begin{proposition}\label{prop:qbp1_info}
Let $G$ be a regular quantum read-once BP and let $Z$ be a classical random variable 
describing an input of~$P$. Then $I(G(Z):Z) = S(G(Z)) \le \log|G|$.
\end{proposition}

The next lemma describes how a regular quantum read-once BP for $\ND_n$ can be used 
for computing the $\AND$ of any pair of input variables $x_i,y_i$ of~$\ND_n$. 

\begin{lemma}\label{lem:restrict}
Let $G$ be a regular $\epsilon$-error quantum read-once BP for $\ND_n$. Let $i\in\{1,\ldots,n\}$ and let
a pair of assignments $(a,b)$ to the variable vectors $(x_j)_{j\neq i}$ and $(y_j)_{j\neq i}$, resp.,
be given such that $\AND(a_j,b_j) = 0$ for all $j\neq i$. 
Then there is an $\epsilon$-error quantum 2-partition protocol~$P_{a,b}$ for $\AND$ on $x_i,y_i$ 
that does not use its public random coin register and has the property
that for each input assignment~$(c,d)$ its result state $P_{a,b}(b,c)$
agrees with the final state $G(a,b,c,d)$ of~$G$ on $(a,b,c,d)$.
\end{lemma}

In the proof of the lemma, we consider quantum read-once BPs with unlabeled nodes obtained by setting
variables to constants as follows. Given a quantum read-once BP~$G$ and a partial assignment~$a$ to some 
of the variables of~$G$, for each node with a variable fixed by~$a$ we remove the variable label,
remove all outgoing edges that are inconsistent with~$a$, and remove all boolean labels
from the remaining edges. Then the resulting graph has the same number of nodes as~$G$ and
still has a well-defined semantics by the remarks in Section~\ref{sec:prelim}.

We prepare the proof of the lemma by describing the modifications of the given quantum BP 
required for the construction of the desired communication protocol in advance. 
Let $G'$ be the quantum read-once BP containing unlabeled nodes obtained from the given graph~$G$ according
to the hypothesis of the lemma by replacing variables with constants according to~$(a,b)$. Observe that due to the fact 
that each variable is read exactly once in~$G$, on each path from the source to a sink in~$G'$ each of the two variables
$x_i$ and~$y_i$ occurs exactly once as the label of a node. Furthermore, each node labeled by a variable
is either the first such node on all paths reaching it or the second. Let $S_x$ and $S_y$ be the
sets of nodes labeled by $x_i$ and $y_i$, resp., where this is the first variable on each 
path reaching the node. Let $S = S_x\cup S_y$. Let $T_x$ and $T_y$ be the sets of immediate successors of nodes
labeled by an $x_i$- or a $y_i$-variable, resp., for which this is the second variable on each path reaching it.
Let $T = T_x\cup T_y$.  To avoid tedious case distinctions, we assume 
w.\,l.\,o.\,g.\ that $S_x\neq\emptyset$ and $S_y\neq\emptyset$. 
In particular, this implies that the source is unlabeled.
Otherwise, it is easy to use the ideas described in the following to define a quantum 1-partition 
communication protocol with the required properties.

Furthermore, for any node~$v$ in~$G'$, let $d_{\rm source}(v)$ and $d_{\rm sinks}(v)$ denote the number of 
edges on each path from the source to~$v$ and the number of edges on each path from~$v$ to a sink, resp. Each of
this is well-defined due to the fact that in the original graph~$G$ each variable occurs
exactly once on each path. The graph~$G'$ has the following properties:

\begin{bulletlist}
\item The sets~$S = S_x\cup S_y$ and~$T = T_x\cup T_y$ form cuts in~$G'$, i.\,e., 
  each path from the source to a sink runs through exactly one node from each set, and all nodes
  on paths from the source to~$S$ (excluding the latter) and from~$T$ to the sinks (including the former)
  are unlabeled nodes.
\item We have $S_x\cap S_y=\emptyset$ and $T_x\cap T_y =\emptyset$ (the latter 
  due to the unidirectionality of~$G'$ inherited from~$G$). Furthermore, all paths starting in~$S_x$
  lead to~$T_y$ and all paths starting in~$S_y$ lead to~$T_x$ and the sets of nodes on the paths of these two types
  are disjoint (for the nodes not in~$S\cup T$, this follows from the read-once property of~$G'$ inherited from~$G$).
\end{bulletlist}

We use these properties to partition the nodes of $G'$ into four subsets:

\begin{bulletlist}\itemsep=4pt
\item The \emph{top part}, including all nodes reached by paths from the source to a node in~$S$,
  excluding the latter;
\item
  two \emph{middle parts} that consist of all nodes on paths from~$S_x$ to~$T_y$ and from~$S_y$ to~$T_x$, resp.; and 
\item a \emph{bottom part} with all nodes on paths starting at a node in~$T$ and 
  leading to a sink, excluding the former and including the latter.
\end{bulletlist}

\goodbreak
Next we further simplify the structure of~$G'$, 
which gives us a new quantum read-once BP with unlabeled nodes. Our aim is to ensure that
all first nodes in the new middle parts lie on a single level and the same 
for all last nodes in the new middle parts.

\medskip
\emph{Changes in the top part:} First, we replace the top part of~$G'$ and extend the 
middle parts upwards. For each node $v\in S$ let $\alpha_v$ be the amplitude for 
reaching it from the source (i.\,e., the sum over all paths of the products of the 
amplitudes at the edges of these paths). Remove the top part of $G'$. 
For each $v\in S$, add a chain of $d_{\rm source}(v)-1 \ge 0$ new unlabeled dummy nodes, 
where a single outgoing edge without boolean label and with amplitude~$1$ 
leads to the next dummy node in the chain for the first $d_{\rm source}(v)-2$ nodes and
to~$v$ for the last dummy node in the chain. Add a new unlabeled source with 
outgoing edges that have no boolean label and that lead to the sources of the 
chains of dummy nodes. The edge leading to the chain of dummy nodes for 
node~$v$ is labeled with amplitude~$\alpha_v$. 

\medskip
\emph{Changes in the bottom part:} 
For a node $v\in T$ and a sink~$w$, let $\beta_{v,w}$ be the amplitude
for reaching~$w$ from~$v$. Remove the bottom part except for the sinks. 
Also remove each node~$v\in T$ and redirect all incoming edges 
to the first node in a chain of new unlabeled dummy nodes of length~$d_{\rm sinks}(v)+1\ge 1$.
The first $d_{\rm sinks}(v)$ nodes of this chain have a single outgoing edge with amplitude~$1$
leading to the next node in the chain. Furthermore, for each sink~$w$ that 
has been reachable from~$v$ in~$G'$ add an edge leading from the last node in the
chain of dummy nodes to~$w$ with amplitude~$\beta_{v,w}$. 
Observe that this construction increases the length of all computation paths by~$1$.
This ensures that the nodes at the ends of the chains of dummy nodes,
which play the role of those in~$T$ in the new graph, are separated from the sinks,
and thus avoids unwanted case distinctions.

\medskip
Call the resulting graph $G''$. 
We state the key properties of~$G''$ in form of the following lemma.

\begin{lemma}\label{lem:dummy}
The graph~$G''$ is a legal quantum read-once BP with unlabeled nodes. Furthermore,
for any input assignment~$(c,d)$ to $(x_i,y_i)$ the final state of~$G''$ on the input~$(c,d)$ 
agrees with that of~$G'$ and thus also with that of~$G$ on the input $(a,b,c,d)$. 
In particular, $G''$ computes the $\AND$ of~$x_i$ and~$y_i$ with the error bound~$\epsilon$ of~$G$. 
\end{lemma}

\begin{proof}
We prove that the changes that turn~$G'$ into~$G''$ retain
well-formedness, unidirectionality, and the transformation computed by the graph as a QBP.
We consider the top part and the bottom part of~$G'$ separately.

\medskip
\emph{Changes in the top part:} We first introduce some notation.
Number the levels of~$G'$ from $0$ (the level of the source) to~$2n$ (the level of the sinks).
Let $L_{\ell}$ be the set of nodes on level~$\ell$. For any subset of nodes $A\subseteq L_{\ell}$ let $\overline{A} = L_{\ell} - A$.
Let $\ell_1 < \cdots < \ell_k$ be the levels of~$G'$ that contain nodes from~$S$ (recall that these
are the first nodes on paths in~$G'$ that are labeled by a variable). Let $\ell_0 = 1 \le \ell_1$.

Observe that the nodes on level~$\ell$ with $0\le\ell\le\ell_k$ can be classified as follows:
(i) nodes in~$S$, i.\,e., nodes that are labeled by a variable and that are reachable only by paths
that solely contain unlabeled nodes;
(ii) unlabeled nodes that are reached only by paths that solely contain unlabeled nodes;
(iii) unlabeled nodes~$v$ with the property that a node in~$S$ lies on each path from the source to~$v$.
Let $V_S(\ell)$, $V_U(\ell)$, resp., be the sets of nodes of the first two types on level~$\ell$.
For levels $\ell,\ell'$ with $\ell < \ell'$ and any sets of nodes $A\subseteq L_{\ell}$ and 
$B \subseteq L_{\ell'}$, where $B$ contains all nodes reachable from~$A$ and no nodes reachable from~$\overline{A}$,
let $U_{A,B}$ be the transformation that acts on the basis vectors of nodes in~$A$ as 
described by the subgraph of~$G'$ consisting of all paths from~$A$ to~$B$ and 
as the identity on all basis vectors of nodes in~$\overline{A}$ (due to the well-formedness
of~$G'$, this can be extended to a unitary transformation). 
For $0\le \ell < \ell'\le 2n$ we use the abbreviation $U_{\ell,\ell'} = U_{L_{\ell},L_{\ell'}}$.

We describe the insertion of the chains of dummy nodes, called \emph{dummy chains}
in the following, as an inductive process. The above definitions of the sets of nodes 
are meant to refer to the actual graph after the modifications carried out so far.
In induction step~$i$, $i=k-1,k-2,\ldots,0$, we modify the levels $\ell_i-1,\ell_i\ldots,\ell_{i+1}$.
The aim is to replace the unlabeled nodes in the top part of~$G'$
between levels $\ell_i$ and $\ell_{i+1}$ by dummy chains and to modify the transformation between levels
$\ell_i-1$ and $\ell_i$ to maintain the correct overall transformation of the graph.

We choose $W'' = V_S(\ell_{i+1})\cup V_U(\ell_{i+1})$ as the set of the end nodes of the new dummy chains.
The set $V_U(\ell_{i+1})$ is empty for $i = k-1$ and contains 
the start nodes of all already constructed dummy chains for $i\le k-2$).
We observe that the nodes on level~$\ell_i$ from which nodes in~$W''$ are reachable are precisely those
in $W' = V_U(\ell_i)$. The immediate predecessors of~$W'$ are the nodes in $W = V_U(\ell_i-1)$.
Furthermore, the set $X = W' \cup V_S(\ell_i)$ contains all nodes reachable from~$W$.
Finally, note that the transformations $U_{W,X}$ and $U_{W',W''}$ are well-defined.

Now step~$i$ of the inductive construction is done as follows:
\begin{bulletlist}
\item Remove all nodes and edges on paths from~$W$ to~$W''$, excluding the start and end nodes. 
\item For each node in~$W''$, insert a chain of new dummy nodes from level~$\ell_i$ to that node.
  Let $\fn{W'}$ denote the set of start nodes of these chains on level~$\ell_i$ and let $T_{W'',\fn{W'}}$ be the 
  linear extension of the bijection 
  that maps the basis states belonging to~$W''$ to those belonging to~$\fn{W'}$.
\item Change the edges between~$W$ and~$X$ such that the transformation $U_{W,X}$ realized before by
  these edges is replaced with the transformation~$T_{W'',\fn{W'}} U_{W',W''}U_{W,X}$.
\end{bulletlist}
Assuming that the given graph is well-formed and unidirectional, these steps can be carried out such that
this is still true for the resulting graph.
We claim that the modifications do not change the transformation realized by the
graph if interpreted as a QBP. We only need to consider the transformations
realized between the modified levels. 

Originally, we have
\[
  U_{\ell_i-1,\ell_{i+1}} \ =\  U_{W',W''} U_{\overline{W'},\overline{W''}}\cdot U_{W,X} U_{\overline{W},\overline{X}}.
\]
Let $\fn{X} = \fn{W'}\cup V_S(\ell_i)$ and denote the transformations in the modified graph between sets~$A$ and~$B$ by $\fn{U}_{A,B}$.
Then, by the construction, $\fn{U}_{W,\fn{X}} = T_{W'',\fn{W'}} U_{W',W''} U_{W,X}$ and $\fn{U}_{\fn{W'},W''} = T_{W'',\fn{W'}}^{-1}$.
Hence, 
\begin{align*}
  \fn{U}_{\ell_i-1,\ell_{i+1}} 
  &\ =\  \fn{U}_{\fn{W'},W''} \fn{U}_{\overline{\fn{W'}},\overline{W''}}\cdot \fn{U}_{W,\fn{X}} \fn{U}_{\overline{W},\overline{\fn{X}}}\\
  &\ =\  T_{W'',\fn{W'}}^{-1} \fn{U}_{\overline{\fn{W'}},\overline{W''}}\cdot T_{W'',\fn{W'}} U_{W',W''} U_{W,X} \fn{U}_{\overline{W},\overline{\fn{X}}}.
\end{align*}
Since $\fn{U}_{\overline{\fn{W'}},\overline{W''}}$ commutes with both $T_{W'',\fn{W'}}^{-1}$ and 
$U_{W',W''}$ due to the disjointness of the respective sets of nodes, we get
\[
  \fn{U}_{\ell_i-1,\ell_{i+1}} 
  \ =\  U_{W',W''} \fn{U}_{\overline{\fn{W'}},\overline{W''}} U_{W,X} \fn{U}_{\overline{W},\overline{\fn{X}}}.
\]
Since the changes do not affect the transformations from~$\overline{W}$ to~$\overline{X} = \overline{\fn{X}}$
and from $\overline{W'} = \overline{\fn{W'}}$ to $\overline{W''}$, the right hand side above is equal to the 
original transformation~$U_{\ell_i-1,\ell_{i+1}}$. Altogether, the graph that we obtain by carrying out all
inductive steps is still well-formed and unidirectional and computes the same transformation as~$G'$. 
It is easy to see that this is exactly the graph obtained by
the modification of the top part described before the lemma.

\medskip
\emph{Changes in the bottom part:} 
For simplicity, we first insert a level of unlabeled dummy nodes directly above the sinks such that each sink
has a corresponding dummy node in this new level which obtains its incoming edges and from which 
it is reached by an edge labeled with amplitude~$1$. This ensures that the set~$T$ of direct successors of nodes that
are the second ones on each path labeled by a variable is disjoint from the sinks in the new graph.
The rest of the proof for the bottom part is now analogous to that for the top part if we look at the
graph turned upside down and exchange the level of the source with that of the sinks and the set~$S$
with the set~$T$.
\end{proof}

We partition the set of nodes of~$G''$ analogously to that of~$G$.
The top part of~$G''$ consists of the source, the bottom part consists
of the sinks, and the two middle parts consist of the nodes in the 
middle parts of~$G'$ together with the dummy nodes on the chains added to the respective parts. 

Overloading notation, we reuse $S,S_x,S_y$ and $T,T_x,T_y$ to denote the sets of start and end nodes, resp.,
in the new middle parts of~$G''$ analogous to the respective sets in~$G'$.
For any node~$v$ and any assignment $z\in\{0,1\}^2$ to~$(x_i,y_i)$, let $\k{\psi_{v,\ell}(z)}$ denote the superposition 
of basis states belonging to the nodes on level~$\ell$ of~$G''$ computed by~$G''$ on input~$z$ 
when starting from the basis state belonging to~$v$.
For the construction of the desired quantum 2-partition protocol, 
we need the following property of~$G''$, which we prove in advance.

\begin{lemma}\label{lem:protorth}
Let $v_1\in S_x$, $v_2\in S_y$ and let $z_1,z_2\in\{0,1\}^2$ be any assignments to $(x_i,y_i)$.
Then for any level~$\ell\in\{1,\ldots,2n+1\}$ (where the source is on level~$0$ and the sinks are
on level~$2n+1$), the states $\k{\psi_{v_1,\ell}(z_1)}$ and 
$\k{\psi_{v_2,\ell}(z_2)}$ are orthogonal.
\end{lemma}

\begin{proof}
The claim is obviously true for all levels $\ell\in\{1,\ldots,2n\}$, since the sets of
nodes in the respective superpositions~$\k{\psi_{v_1,\ell}(z_1)}$ and~$\k{\psi_{v_2,\ell}(z_2)}$
are disjoint. We have to verify the claim for the level~$\ell = 2n+1$ of the sinks.

Let~$U$ be a unitary extension of the transformation realized by the edges between
the last level~$2n$ of~$G''$ above the sinks and level~$2n+1$. Since the last level
above the sinks only contains unlabeled nodes, $U$ does not depend on the input.
Thus, $\k{\psi_{v_i,2n+1}(z_i)} = U\k{\psi_{v_i,2n}(z_i)}$ for $i=1,2$ and
we get
\belowdisplayskip=0pt
\begin{align*}
  \bk{\psi_{v_1,2n+1}(z_1)}{\psi_{v_2,2n+1}(z_2)}
  &\ =\  \b{\psi_{v_1,2n}(z_1)}U^\dag U\k{\psi_{v_2,2n}(z_2)}\\
  &\ =\  \bk{\psi_{v_1,2n}(z_1)}{\psi_{v_2,2n}(z_2)} 
  \ =\  0.
\end{align*}
\end{proof}

\begin{proof}[Proof of Lemma~\ref{lem:restrict}]
We construct the quantum $2$-partition protocol~$P = P_{a,b}$ using the graph~$G''$.
We make sure that~$P$ simulates~$G''$.

For $v\in S = S_x\cup S_y$, let $\alpha_v$ be
the amplitude for reaching~$v$ from the source of~$G''$.
The protocol~$P$ works on the space spanned 
by the basis vectors belonging to the nodes in the middle parts and in the bottom part of~$G''$. 
It has subprotocols $P_x$ and $P_y$ with respect to the variable partitions $(\{x_i\},\{y_i\})$ and
$(\{y_i\},\{x_i\})$, resp. Let $q_x = \sum_{v\in S_x} |\alpha_v|^2$ and
$q_y = \sum_{v\in S_y} |\alpha_v|^2$.
The initial amplitudes of $P_x$ and $P_y$ are defined as $\sqrt{q_x}$ and $\sqrt{q_y}$, resp.
As the initial state~$\k{s_x}$ of $P_x$ we choose $\sum_{v\in S_x} (\alpha_v/\sqrt{q_x})\k{v}$ 
if $q_x \neq 0$ and some arbitrary $\k{v}$ with $v\in S_x$ if $q_x = 0$. 
Define $\k{s_y}$ analogously for~$P_y$. 

We only describe the computation of $P_x$ in detail, $P_y$ works in the same way.
We first define further subprotocols $P_{x,v}$ belonging to each of the nodes $v\in S_x$.
Let $G_v''$ be the subgraph of $G''$ with source $v\in S_x$ containing all nodes 
reachable from~$v$. On each path starting at a node~$v\in S_x$ there 
is exactly one $y_i$-node. There is some level~$m$ where the first $y_i$-node 
in~$G_v''$ is read. In $P_{x,v}$ the player Alice simulates the 
computation of $G_v''$ starting at~$v$ and until level~$m$. 
She sends the reached superposition of basis states of nodes on 
level~$m$ to Bob. Bob continues the simulation of~$G_v''$ starting
with the superposition received from Alice and computing a superposition of the sinks in~$G_v''$. 
Let $P_x$ be the protocol where the described subprotocols $P_{x,v}$, $v\in S_x$, are applied to the 
initial state of~$P_x$. 

We claim that $P_x$ designed in this way is a legal quantum one-way protocol.
The state obtained after Alice has finished her computation in~$P_x$ 
need not be reachable by any computation in~$G''$. Nevertheless, it is a legal pure quantum state
by the following argument. Let $S_{x,m}\subseteq S_x$ be the set of all nodes~$v$ for which~$m$ is the 
first level with $y_i$-nodes reached from~$v$ in~$G_v''$. Let~$A_v$ be the unitary transformation 
applied by Alice in~$P_{x,v}$. Then the state computed by Alice according to $P_x$ is
$\k{\psi} = \sum_{v\in S_x} \alpha_v A_v\k{v}$ and we have
\begin{align*}
  \bk{\psi}{\psi} 
  & =\  \sum_{v,v'\in S_x} \alpha_v^*\alpha_{v'} \b{v}A_v^\dag A_{v'}\k{v'}
  \ =\  \sum_{m} \sum_{v,v'\in S_{x,m}} \alpha_v^*\alpha_{v'} \b{v}A_v^\dag A_{v'}\k{v'}\\
  &\ =\  \sum_{m} \sum_{v\in S_{x,m}} \bigl|\alpha_v\bigl|^2
  \ =\  1.
\end{align*}
The second equality is due to the fact that the subspaces induced by the nodes on different
levels of~$G''$ are orthogonal. The third equality follows from the unitarity of the time
evolution of~$G''$. Due to the same fact, also Bob's transformation 
in $P_x$ is unitary.

Finally, due to Lemma~\ref{lem:protorth}, the state spaces of $P_x$ and~$P_y$ constructed
in the above way are orthogonal. Hence, putting these protocols together as described 
before gives a legal quantum $2$-partition protocol~$P$. 
It is obvious that $P$ simulates $G''$ and thus its
result state also agrees with the final state of~$G$.
\end{proof}

We are now ready to prove the main theorem.

\begin{proof}[Proof of Theorem~\ref{the:disj}]
{\sloppy\hbadness=10000
Let $G$ be a regular $\epsilon$-error quantum read-once BP for~$\ND_n$. 
We run~$G$ on the random input $\vec{Z}$ conditioned on $\vec{D} = \vec{d}$.
Since $\vec{Z} = (Z_1,\ldots,Z_n)$, where $Z_1,\ldots,Z_n$ are 
independent, Fact~\ref{fact:info}(\ref{facti:concavity_info}) (superadditivity of mutual information), 
Fact~\ref{fact:info}(\ref{facti:convex_info}), and Proposition~\ref{prop:qbp1_info} yield
\begin{align*}
  \sum_{i=1}^n I(G(\vec{Z})\,{:}\,Z_i\,|\,\vec{D}) 
  \,\le\, I(G(\vec{Z})\,{:}\,\vec{Z}\,|\,\vec{D})
  \,=\, \!{\sum_{\vec{d}} \Pr\{ \vec{D} = \vec{d} \}\cdot I(G(\vec{Z})\,{:}\,Z\,|\,\vec{D}=\vec{d})}
  \,\le\,  \log |G|.
\end{align*}
Hence, by averaging, we can fix an~$i$ such that 
\[
  I(G(\vec{Z})\,{:}\,Z_i\,|\,\vec{D}) \ \le\  (\log|G|)/n.
\]
Let $\vec{D} = (\vec{D}_{-i},D_i)$, where $\vec{D}_{-i} = (D_j)_{j\neq i}$. 
Then again by averaging and Fact~\ref{fact:info}(\ref{facti:convex_info}), 
there is a value $\vec{d}_{-i}$ for $\vec{D}_{-i}$ such that
\[
  I(G(\vec{Z})\,{:}\,Z_i\,|\,\vec{D}_{-i} = \vec{d}_{-i}, D_i) 
  \ \le\  (\log|G|)/n.
\]
To prove the claim, we lower bound the term on the left hand side of this inequality by the 
information cost of an $\epsilon$-error quantum 
2-partition protocol for $\AND$ on input~$Z_i$ conditioned on~$D_i$.
Then using the constant lower bound on the information cost from Theorem~\ref{the:and} 
and the above inequality, we get that $(\log|G|)/n = \Omega(1)$ and thus $|G| = 2^{\Omega(n)}$, 
which proves the theorem.

}

For the following, let $d$ be any fixed value for~$D_i$.
Let $\vec{Z}^{(d)} = \bigl(Z_1^{(d)},\ldots,Z_n^{(d)}\bigr)$ 
be a random variable that is distributed as~$\vec{Z}$
conditioned on $\vec{D}_{-i} = \vec{d}_{-i}$ and $D_i = d$. 
Let $\vec{Z}^{(d)}_{-i} = \bigl(Z_j^{(d)}\bigr)_{j\neq i}$.
For each fixed value~$\vec{z}_{-i}$ in the support of $\vec{Z}^{(d)}_{-i}$,
we get an $\epsilon$-error quantum 2-partition protocol~$P_{\vec{z}_{-i}}$ 
for~$\AND$ on the input~$z_i = (x_i,y_i)$ by Lemma~\ref{lem:restrict}. 
This protocol does not use its public random coin register.
Furthermore, the result state of~$P_{\vec{z}_{-i}}$ on~$z_i$ agrees
with the final state $G(\vec{z}_{-i},z_i)$ of~$G$.
Let $Q$ be a quantum 2-partition protocol in which the players run~$P_{\vec{z}_{-i}}$ 
for $\vec{z}_{-i}$ chosen randomly with the distribution of~$\vec{Z}_{-i}$ under
the condition $\vec{D}_{-i} = \vec{d}_{-i}$. They can do this by initializing
the public random coin register and the secret part of this register appropriately
according to our conventions.
Then the result state of~$Q$ after trace-out of the input registers
of both players, the random coin register, and the secret registers
is $Q(z_i) = P_{\vec{Z}^{(d)}_{-i}}(z_i)$.

Now $Q$ is run on the random input~$Z_i^{(d)}$ by using the
secret input registers of Alice and Bob (at this point, we exploit the fact that 
the input bits of Alice and Bob under the condition~$D_i = d$ are independent of each other). 
Expanding the abbreviations and using the fact that
$P_{\vec{z}_{-i}}(z_i) = G(\vec{z}_{-i},z_i)$ for all $(\vec{z}_{-i},z_i)$, we get:
\begin{align*}
  I\bigl(Q\bigl(Z_i^{(d)}\bigr)\,{:}\,Z_i^{(d)}\bigr) 
  \ =\  I(Q(Z_i) \,{:}\, Z_i \,|\, D_i = d )
  &\ =\  I\bigl(P_{\vec{Z}_{-i}}\bigl(Z_i\bigr) \,{:}\, Z_i \,|\, \vec{D}_{-i} = \vec{d}_{-i}, D_i=d\bigr)\\
  &\ =\  I\bigl(G\bigl(\vec{Z}\bigr) \,{:}\, Z_i \,|\, \vec{D}_{-i} = \vec{d}_{-i}, D_i=d\bigr).
\end{align*}
Averaging over all values~$d$ yields 
\[
  \IC(Q;Z_i\,|\,D_i) \ =\  I(Q(Z_i) \,{:}\, Z_i \,|\, D_i) 
  \ =\  I\bigl(G\bigl(\vec{Z}\bigr) \,{:}\, Z_i \,|\, \vec{D}_{-i} = \vec{d}_{-i}, D_i\bigr).
\]
Since for the vector $\vec{Z}'$ obtained from~$\vec{Z}$ by replacing $Z_i$ with any $z_i\in\{0,1\}^2$,
$\ND_n(\vec{Z}') = \AND(z_i)$ with probability~$1$, we know that~$Q$ is an $\epsilon$-error
quantum 2-partition protocol for~$\AND$. By the lower bound on the information
cost of quantum multi-partition protocols for~$\AND$ from Theorem~\ref{the:and}, 
it follows that the left hand side of the above inequality is lower bounded
by a positive constant.
Together with our above arguments, this completes the proof.
\end{proof}

\section*{Acknowledgment}

The author wishes to thank Detlef Sieling for proofreading of draft versions of the 
present paper, several valuable suggestions regarding proof details and presentation, 
and for a lot of discussions about quantum branching programs in general. 
Furthermore, the helpful comments of several anonymous referees are 
gratefully acknowledged.

\hbadness=10000

\end{document}